\documentclass[amsmath,amssymb,12pt,superscriptaddress,nofootinbib]{revtex4-1}

\usepackage{graphicx}
\usepackage{dcolumn}
\usepackage{bm}
\usepackage{color}
\usepackage{enumitem}
  \usepackage{tabularx}
   \newcolumntype{C}{>{\centering\arraybackslash}X}
   \newcolumntype{L}{>{\raggedright\arraybackslash}X}
   \newcolumntype{R}{>{\raggedleft\arraybackslash}X}
\setlistdepth{10}

\newcommand{\dd}{\mathrm{d}}

\newcommand{\del}{\partial}
\newcommand{\ee}{{\rm e}}

\definecolor{DarkBlue}{rgb}{0,0,0.7} 

\definecolor{DarkRed}{rgb}{0.65,0,0} 
\newcommand{\dred}[1]{{#1}}

\begin{document}
\baselineskip5.5mm


{\baselineskip0pt
\small
\leftline{\baselineskip16pt\sl\vbox to0pt{
                             \vss}}
\rightline{\baselineskip16pt\rm\vbox to20pt{
\vspace{-1.5cm}
            \hbox{RUP-18-15}
            \hbox{KEK-Cosmo-225}
            \hbox{KEK-TH-2052}
\vss}}
}

\author{Chul-Moon~Yoo}\email{yoo@gravity.phys.nagoya-u.ac.jp}
\affiliation{
\fontsize{11pt}{0pt}\selectfont
Division of Particle and Astrophysical Science,
Graduate School of Science, Nagoya University, 
Nagoya 464-8602, Japan
\vspace{1.5mm}
}

\author{Tomohiro~Harada}\email{harada@rikkyo.ac.jp}
\affiliation{~
\fontsize{11pt}{0pt}\selectfont
Department of Physics, Rikkyo University, Toshima,
Tokyo 171-8501, Japan
\vspace{1.5mm}
}

\author{Jaume~Garriga}\email{jaume.garriga@ub.edu}
\affiliation{
\fontsize{11pt}{0pt}\selectfont
Departament de F\'isica Qu\`antica i Astrof\'\i sica, Institut de Ci\`encies del Cosmos, Universitat de
Barcelona, Mart\'\i\ i Franqu\`es 1, 08028 Barcelona, Spain
\vspace{1.5mm}
}

\author{Kazunori~Kohri}\email{kohri@post.kek.jp}
\affiliation{
\fontsize{11pt}{0pt}\selectfont
Rudolf Peierls Centre for Theoretical Physics, The University of
Oxford, 1 Keble Road, Oxford, OX1 3NP, UK
\vspace{1.5mm}
}
\affiliation{
\fontsize{11pt}{0pt}\selectfont
Institute of Particle and Nuclear Studies, KEK, 1-1 Oho, Tsukuba, Ibaraki 305-0801, Japan
\vspace{1.5mm}
}
\affiliation{
\fontsize{11pt}{0pt}\selectfont
The Graduate University for Advanced Studies (SOKENDAI), 1-1 Oho, Tsukuba, Ibaraki 305-0801, Japan
\vspace{1mm}
}


\vskip-1.5cm
\title{PBH abundance from random Gaussian curvature perturbations 
and a local density threshold}


\begin{abstract}
\baselineskip5mm 
\vskip-0.5cm 
The production rate of primordial black holes is often calculated by considering 
a nearly Gaussian distribution of cosmological perturbations, and assuming
that black holes will form in regions where the amplitude of such perturbations exceeds a certain threshold. 
A threshold $\zeta_{\rm th}$ for the curvature perturbation is somewhat inappropriate for this purpose, because it 
depends significantly on environmental effects, not essential to the local dynamics.  
By contrast, a threshold $\delta_{\rm th}$ for the density perturbation at horizon crossing 
seems to provide a more robust criterion. 
On the other hand, the density perturbation is known to be
bounded above by a maximum limit $\delta_{\rm max}$ at the horizon entry, 
and given that $\delta_{\rm th}$ is 
comparable to $\delta_{\rm max}$,
the density perturbation will be far from Gaussian near or above the threshold.
In this paper, we provide a new plausible estimate for the primordial black hole abundance based on peak theory.  
In our approach, we assume that the curvature perturbation is given as a random Gaussian field 
with the power spectrum characterized by a single scale, 
while an optimized criterion for PBH formation is imposed, based on the locally averaged density perturbation 
around the nearly spherically symmetric high peaks. 
Both variables are related by the full nonlinear expression derived in 
the long-wavelength approximation of general relativity. 
We do not introduce a window function which is usually introduced to obtain the scale dependence of the spectrum. 
The scale of the inhomogeneity is introduced as a random variable in the peak theory, 
and the scale dependent PBH fraction is automatically induced. 
We find that the mass spectrum is shifted to larger mass scales by one order of magnitude or so, 
compared to a conventional calculation. 
The abundance of PBHs 
becomes significantly larger than the conventional one, by many orders of magnitude,
mainly due to the optimized criterion for PBH formation 
and the removal of the suppression associated with a window function.
\end{abstract}


\maketitle
\thispagestyle{empty}
\pagebreak

\section{Introduction}
\label{sec:intro}

Processes which may lead to the formation of primordial black holes (PBHs), 
along with their cosmological implications, 
have been extensively investigated in the literature since the pioneering work of Zel'dovich and Novikov~\cite{1967SvA....10..602Z} and
Hawking~\cite{Hawking:1971ei}.  PBHs may be produced by gravitational collapse
in regions with a large amplitude of density perturbations in the early Universe, and measurements or constraints on their
abundance can be regarded as a probe of the primordial spectrum and the underlying inflationary model.
The latest observational constraints are summarized in,
e.g. Refs.~\cite{Carr:2009jm,Carr:2016drx}.  
So far, PBHs have been actively studied as candidates of dark matter (e.g., see
Refs.~\cite{1975Natur.253..251C,Carr:1975qj,GarciaBellido:1996qt,Jedamzik:1999am,Frampton:2010sw,Kawasaki:2012wr,Kohri:2012yw,Axelrod:2016nkp,Carr:2016drx,Ezquiaga:2017fvi,Clesse:2017bsw,Kohri:2018qtx}
and references therein).  In addition, the recent discovery of gravitational waves emitted from binary black holes
(BBHs)~\cite{Abbott:2016blz,Abbott:2017vtc} has stimulated the investigation of PBH binaries and their merger rates~\cite{Sasaki:2016jop,Bird:2016dcv,Clesse:2016vqa,Raidal:2017mfl}.

In this paper, we will focus on the formation of PBHs in the radiation dominated era (see
Refs.~\cite{Khlopov:1980mg,Harada:2016mhb,Harada:2017fjm} for cases of
the matter dominated era), due to some enhanced feature in the primordial power spectrum of density perturbations around some specific scale.\footnote{ \baselineskip 4.5mm 
Inflation can also produce relics, such as vacuum bubbles and domain walls with a distribution of sizes, which may in turn produce PBH during the subsequent radiation dominated era (see e.g. \cite{Tanahashi:2014sma,Garriga:2015fdk,Deng:2018cxb} and references therein). In this case, the relics behave as active seeds, and trigger gravitational collapse with unit probability if their initial comoving size is sufficiently large. Our present considerations do not apply to such situation.}
A rough criterion for the formation of PBHs was first proposed by Carr~\cite{Carr:1975qj}, and much numerical
work has been done in search of a more accurate criterion~\cite{1978SvA....22..129N,1980SvA....24..147N,Shibata:1999zs,Niemeyer:1999ak,Musco:2004ak,Polnarev:2006aa,Musco:2008hv,Musco:2012au,Polnarev:2012bi,Nakama:2013ica,Nakama:2014fra}.  The perturbation variables which are used to characterize the amplitude of
the initial inhomogeneity are roughly divided into two sorts: the
density perturbation and the curvature perturbation.  For instance,
Shibata and Sasaki~\cite{Shibata:1999zs} discussed the threshold for PBH formation by using the curvature variable which is given by
the conformal factor of the spatial metric. On the other hand, in
Refs.~\cite{Musco:2004ak,Polnarev:2006aa,Musco:2012au,Polnarev:2012bi,Nakama:2013ica,Nakama:2014fra},
the threshold value is given for the averaged density perturbation at
the horizon entry in the comoving slicing, and in the lowest order
long-wavelength expansion.  The threshold value of the density
perturbation is given by $\delta_{\rm th}\approx 0.42 - 0.66$ depending on the perturbation
profile.  The lowest threshold value seems to correspond to the value analytically derived in
Ref.~\cite{Harada:2013epa} with significant simplification~\cite{Musco2018}.

As for the curvature variable, it has been suggested in
Ref.~\cite{Young:2014ana} that the threshold is strongly affected by
environmental effects, while that of a density perturbation is not.
This fact has been also numerically demonstrated in
Ref.~\cite{Harada:2015yda}.  One extreme example which shows the
significance of the environmental effects is the estimate of the threshold of the curvature
perturbation suggested in Ref.~\cite{Harada:2013epa}.  There, an (irrelevant) extremely low value of
$\zeta_{\rm th}\simeq 0.0862$ is obtained, due to the existence of an unphysical negative
density region in the environment in the specific model adopted there.
However, even if we keep the positivity of the density in the
environment, the threshold value of the curvature perturbation can be
significantly affected~\cite{Harada:2015yda}.  This can be intuitively
understood if we consider the curvature perturbation as the general
relativistic counterpart of the Newtonian potential, which can be shifted by an arbitrary constant.  Since, at
least in spherically symmetric systems, the process of PBH formation can be
described in a quasi-local manner, the use of a threshold value for a
quasi-local perturbation variable seems to be more appropriate (see
Sec.VII and VIII in Ref.~\cite{Harada:2015yda} for details).

A useful criterion has been proposed in Ref.~\cite{Shibata:1999zs} for spherically symmetric systems 
based on the so-called compaction function, which is 
equivalent to one half of the volume average of the density perturbation $\delta$ 
at the time of horizon entry \cite{Harada:2015yda}.
The criterion for PBH formation is that the maximum value 
of the compaction function as a function of the averaging radius
lies above a specified threshold $\mathcal C_{\rm th}=\delta_{\rm th}/2$, 
at the time when the averaging radius enters the horizon.  
Such threshold value has been found to be 
more
robust. This has been tested by considering two different families of profiles for the perturbation, and a broad range of parameters~\cite{Harada:2015yda}. 
In what follows, we will not further
discuss the possible profile dependence of the threshold, but simply assume the
existence of a typical value(see e.g. Ref.~\cite{Musco2018} for an analysis about the profile dependence). 
We also note that, although our framework is applicable to generic non-spherical systems, 
we will adopt a criterion for PBH formation by referring to the compaction function in the corresponding spherical 
system. This is justified because high peaks of a random Gaussian field tend to be spherical.

The main purpose of this paper is to find an estimate for the abundance of PBHs
once a threshold value of 
the averaged density perturbation is provided.  
One
conventional way is to apply the Press-Schechter(PS) formalism to the
density perturbation by assuming that this variable is Gaussian distributed. However, 
due to the local dynamics of overdense regions, 
there is an upper limit for the value of the density perturbation at horizon crossing.
This was first observed in Refs.~\cite{Kopp:2010sh,Harada:2013epa}, in the context of
a simplified ``3 zone model" where a spherical homogeneous overdensity
is embedded in a flat Friedmann-Lema\^itre-Robertson-Walker(FLRW) environment.  
More generally, it was found  \cite{Harada:2015yda} 
that for spherically symmetric perturbations with any profile, the maximum density
perturbation at horizon crossing in the co-moving slicing is bounded by $\delta_{\max} \approx 2/3$. 
The argument will be reviewed in Section \ref{sec2}.\footnote{
 \baselineskip 4.5mm 
In Ref.~\cite{Kopp:2010sh} the maximum 
density perturbation at horizon crossing in the geodesic slicing  is found to be $\delta^G_{\rm max} = 9/16$. 
In the long wavelength approximation \cite{Harada:2015yda}, such value translates into the comoving slicing as
$\delta_{\rm max} \approx 3/4$. Note that this differs from the determination given in \cite{Harada:2013epa,Harada:2015yda} by a factor of $8/9$,
which may be related to the extrapolation of the long wavelength approximation in relating the different gauges at horizon crossing.}
Noting that $\delta_{\rm th}$ is in fact comparable to the maximum value 
$\delta_{\rm max}$ (above which the probability distribution should vanish) a naive application of the Gaussian distribution seems rather questionable.
In addition, while the threshold is often set for the density
perturbation, the statistical properties of the primordial curvature perturbation are
usually better understood.  Therefore, it is
natural to consider the abundance of PBHs by combining the threshold of
the density perturbation with the statistical properties of the
curvature perturbation. Since PBH formation is a non-linear
process, a non-linear relation between these perturbation variables
should be taken into account.  In this paper, we address the calculation of the PBH abundance
by using the peak theory for the Gaussian probability
distribution of the curvature perturbation, and the non-linear relation between
curvature and density perturbation in the long-wavelength limit. 
Readers not interested in the details of the derivation can skip
directly to Eq.~\eqref{eq:beta_general}, and the 
ensuing explanation on how to use it.

Another significant problem is the window function dependence of the mass spectrum. 
In Ref.~\cite{Ando:2018qdb}, it is reported that the mass spectrum significantly depends on the choice of the window function in the PS formalism. 
For an extended curvature power spectrum, the scale dependence of the PBH fraction is introduced by a window function 
in the PS formalism. 
Therefore, without a window function, an extended mass spectrum of PBH cannot be obtained along the conventional PS formalism. 
In contrast, according to the peak theory, the scale of the inhomogeneity can be also introduced as a random variable. 
The probability distribution of the random variable is associated 
with the power spectrum of the curvature perturbation. 
Then, in our procedure, the scale dependence is naturally induced depending on the profile of the curvature power spectrum.

The plan of the paper is the following. 
In Section II we discuss the perturbation variables and the implementation of a threshold for PBH formation.
In Section III we consider the statistical properties of the Taylor expansion coefficients of the Gaussian random field $\zeta$ around a given point. 
In Section IV, where we
discuss the probability for PBH formation, 
based on the peak theory and on our implementation of the threshold on the averaged density perturbation. 
The results will be compared to the more conventional PS approach for the monochromatic and a simple extended 
curvature power spectrum in Section V. 
Our conclusions are summarized in Section V.
Some technical aspects are discussed in the Appendices. 
Throughout this paper, we use the geometrized units in which both
the speed of light and Newton's gravitational constant are unity, $c=G=1$.

\section{perturbation variables and threshold for PBH formation}
\label{sec2}

Let us consider the 
density perturbation in the comoving slicing, which is
orthogonal to the fluid world line. In the long-wavelength
approximation, the
curvature perturbation $\zeta$ and the density perturbation $\delta$
with the comoving slicing are related by~\cite{Harada:2015yda},
\begin{equation}
\delta=-\frac{4(1+w)}{3w+5}\frac{1}{a^2H^2}\ee ^{5\zeta/2}\triangle\left(\ee^{-\zeta/2}\right), 
\end{equation}
where $w$ is the equation of state parameter, $a$ is the scale factor, $H$ is the Hubble rate, 
$\triangle$ is the Laplacian of the reference flat metric, 
and the spatial metric is given by 
\begin{equation}
\dd s_3^2=a^2\ee^{-2\zeta}\tilde \gamma_{ij}\dd x^i \dd x^j, 
\end{equation}
with $\det \tilde \gamma $ being the same as the determinant of the reference flat metric. 

We will be interested mainly in high peaks, 
which tend to be nearly spherically symmetric \cite{1986ApJ...304...15B}.
Therefore, in this section, we introduce the criterion for PBH formation 
assuming spherical symmetry originally proposed in Ref.~\cite{Shibata:1999zs}. 
Here, we basically follow and refer to the discussions and calculation in Ref.~\cite{Harada:2015yda}. 

First, let us define the compaction function $\mathcal C$ as 
\begin{equation}
\mathcal C:=\frac{\delta M}{R}, 
\end{equation}
where $R$ is the areal radius at the radius $r$, 
and $\delta M$ is the excess of the Misner-Sharp mass $M_{\rm MS}$ enclosed by the sphere of the radius $r$ compared with the 
mass $M_{\rm F}$ inside the sphere in the fiducial flat FLRW universe with the same areal radius. 
The Misner-Sharp mass for the comoving slicing is given by 
\begin{equation}
M_{\rm MS}(r)=4\pi\int^r_0\dd x \rho R^2 R', 
\end{equation}
where $\rho$ is the matter density and the prime ``$~'~$" denotes the derivative with respect to the radial coordinate(see Sec.~V-B in Ref.~\cite{Harada:2015yda} for general slicing). 
Then, 
we find the following expressions for each variable:
\begin{eqnarray}
M_{\rm MS}(r)&=&\frac{3}{2}H^2\int^r_0\dd x x^2(1+\delta)\ee^{-3\zeta}\left(1-r\zeta'\right), \\
M_{\rm F}(r)&=&\frac{1}{2}H^2a^3r^3, \\
\mathcal C(r)&=&\frac{M_{\rm MS}(r)-M_{\rm F}(r\ee^{-\zeta})}{ar\ee^{-\zeta}}. 
\end{eqnarray}
In the limit of the long-wavelength approximation, for comoving and constant mean curvature gauge, 
the compaction function is given by: 
\begin{equation}
\mathcal C=\frac{1}{2}\bar \delta \left(HR\right)^2, \label{onehalf}
\end{equation}
where $\bar \delta$ is the volume average of the density perturbation $\delta$ within the radius $r$ (see Eq.~(5.31) in Ref.~\cite{Harada:2015yda}). 
From the definition of $\mathcal C$, we can derive the following simple form in the comoving slicing (see also Eq.~(6.33) in Ref.~\cite{Harada:2015yda}):
\begin{equation}
\mathcal C(r)=\frac{1}{3}\left[1-\left(1-r\zeta'\right)^2\right]. 
\label{eq:Compaction}
\end{equation}
From this expression, it is clear that $\mathcal C \leq1/3.$ If we identify the time of horizon entry of the perturbation from the condition 
$HR=1$, then, form Eq. (\ref{onehalf}), the corresponding averaged density perturbation is $\bar \delta < \delta_{\rm max} = 2/3$, as discussed in the introduction. 
We can rewrite $\mathcal C(r)$ as $\left(1-R'^2\ee^{2\zeta}/a^2 \right)/3$. 
The existence of the region $R'<0$ corresponds to the Type II PBH formation reported in Ref.~\cite{Kopp:2010sh}. 
In what follows, we focus on the Type I cases, that is, $R'>0$.

We will also assume that the function $\mathcal C$ is a smooth function of $r$ for $r>0$. 
Then, the value of $\mathcal C$ takes the maximum value $\mathcal C^{\rm max}$ at $r_{\rm m}$ 
which satisfies\footnote{
\baselineskip 4.5mm
We thank I.~Musco for pointing out the importance of the 
radius $r_{\rm m}$ to us \cite{priv_Ilia}.
}
\begin{equation}
\mathcal C'(r_{\rm m})=0\Leftrightarrow \left(\zeta'+r\zeta''\right)|_{r=r_{\rm m}}=0. 
\label{eq:forrm}
\end{equation}
We consider the following criterion for PBH formation:
\begin{equation}
\mathcal C^{\rm max}>\mathcal C_{\rm th} \equiv \frac{1}{2}\delta_{\rm th}. \label{maxth}
\end{equation}
In the constant-mean-curvature(CMC) slice, the threshold $\mathcal C^{\rm CMC}_{\rm th}$ for PBH formation is evaluated as $\simeq0.4\pm0.03$ (see Figs.~2 and 3 or TABLE I and II in Ref.~\cite{Harada:2015yda}). 
This threshold corresponds to the perturbation profiles of Refs.~\cite{Shibata:1999zs,Polnarev:2006aa,Musco:2012au}, 
and is found to be quite robust for a broad range of parameters. 
Since the relation between the density perturbation in the comoving slice ($\delta$) and the CMC slice ($\delta_{\rm CMC}$) is given by 
\begin{equation}
\delta=\frac{2}{3}\delta_{\rm CMC}, 
\end{equation}
the threshold value in the comoving slice is 
given by $\mathcal C_{\rm th}\simeq0.267$ which corresponds to $\delta_{\rm th}\simeq 0.533$. 
In this paper we shall use this as a reference value. 
We should keep in mind, however, that the threshold value is not completely 
independent of profile. 
For instance, as mentioned in the introduction, 
the threshold in a 3-zone model with a homogeneous overdensity the 
threshold is somewhat lower.

Throughout this paper, we assume the random Gaussian distribution of $\zeta$ with
its power spectrum $\mathcal P(k)$ defined by the following equation:
\begin{equation}
<\tilde \zeta^*(\bm k)\tilde \zeta(\bm k')>=\frac{2\pi^2}{k^3}\mathcal P(k)(2\pi)^3\delta(\bm k-\bm k'), 
\end{equation}
where $\tilde \zeta(\bm k)$ is the Fourier transform of $\zeta$ and 
the bracket $<...>$ denotes the ensemble average. 
Each gradient moment $\sigma_n$ can be calculated by 
\begin{equation}
\sigma_n^2:=\int \frac{\dd k}{k}k^{2n} \mathcal  P(k). 
\end{equation}

Focusing on a high peak and taking it as the origin of the coordinates, we introduce the amplitude $\mu$ and the curvature scale $1/k_*$ of the peak as follows: 
\begin{eqnarray}
\mu&=&-\left.\zeta\right|_{r=0}, 
\label{eq:mudef}\\
k_*^2&=&\frac{\left.\triangle \zeta\right|_{r=0}}{\mu}. 
\label{eq:ksdef}
\end{eqnarray}
Peaks of $\zeta$ do not necessarily correspond to peaks of $\delta$. 
Nevertheless, as is shown in Appendix~\ref{devpeaks}, 
if the value of $\delta$ is
comparable to the threshold value $\delta_{\rm th}$ at a peak, we can
almost always find the associated peak of $\zeta$ well inside the
horizon patch centered at the peak of $\delta$. 
In some cases, there may be a local minimum of $\delta$ where there is a 
maximum of $\zeta$, although this typically happens for values of $\zeta$ 
above the threshold for black hole formation. 
This is discussed in 
Appendix~\ref{sec:k-relation}, where we also give the relation between $k_*$ and 
the inverse length scale $k_\delta \equiv (-\triangle \delta/\delta|_{{\bf x}=0})^{1/2}$ of the density perturbation at the peak.
According to the peak theory\cite{1986ApJ...304...15B}, 
for a high peak, we may expect the typical form of the profile $\bar \zeta$ can be described by using 
$\mu$, $k_*$ and the two point correlation function $\psi$ as follows:
\begin{equation}
\frac{\bar \zeta(r)}{\mu}=g(r;k_*):=
g_0(r)+k_*^2g_1(r), 
\label{eq:zetahat}
\end{equation}
where 
\begin{eqnarray}
g_0(r)&=&-\frac{1}{1-\gamma^2}\left(\psi+\frac{1}{3}R_*^2\triangle \psi\right),
\label{eq:g0}
\\  
g_1(r)&=&\frac{ 1}{\gamma(1-\gamma^2)}\frac{\sigma_0}{\sigma_2}
\left(\gamma^2\psi+\frac{1}{3}R_*^2\triangle \psi\right). 
\label{eq:g1}
\end{eqnarray}
with $\gamma=\sigma_1^2/(\sigma_0\sigma_2)$, $R_*=\sqrt{3}\sigma_1/\sigma_2$ and 
\begin{equation}
\psi(r)=\frac{1}{\sigma_0^2}\int\frac{\dd k}{k}\frac{\sin(kr)}{kr}\mathcal P(k). 
\label{eq:psi}
\end{equation}
It is worthy of note that, for $k_*=k_{\rm c}:=\sigma_1/\sigma_0$, 
we obtain
\begin{equation}
g(r;k_{\rm c})=-\psi(r). 
\label{eq:gkc}
\end{equation}
It will be shown that 
regarding $k_*$ as a probability variable, we obtain $k_{\rm c}$ as the mean value of $k_*$. 
The profile \eqref{eq:gkc} is introduced in the recent paper \cite{Germani:2018jgr} as a typical profile associated with the curvature power spectrum. 
Here, we also take $k_*$ dependence into account, and introduce the scale dependence to the profile.

Applying Eq.~\eqref{eq:Compaction} to $\bar\zeta$, we obtain the relation between $\mu$ and $\mathcal C$ as 
\begin{equation}
\mu=\frac{1-\sqrt{1-3\mathcal C}}{rg'}, 
\end{equation}
where the smaller root is taken. 
Let us define the threshold value $\mu_{\rm th}^{(k_*)}$ as 
\begin{equation}
\mu_{\rm th}^{(k_*)}(k_*)=\frac{1-\sqrt{1-3\mathcal C_{\rm th}}}{\bar r_{\rm m}(k_*)g_{\rm m}'(k_*)}=\frac{2-\sqrt{4-6\delta_{\rm th}}}{2\bar r_{\rm m}(k_*)g_{\rm m}'(k_*)}, 
\label{eq:muth}
\end{equation}
where 
$\bar r_{\rm m}(k_*)$ is the value of $r_{\rm m}$ for $\zeta=\bar \zeta$, and
\begin{equation}
g_{\rm m}(k_*):=g(\bar r_{\rm m}(k_*);k_*). 
\end{equation}
In Eq.~\eqref{eq:muth}, we have explicitly denoted the $k_*$ dependence of $\bar r_{\rm m}$ and $g_{\rm m}$ to emphasize it. 

Although we obtain the threshold of the amplitude $\mu$ as a function of $k_*$, 
since our goal is to obtain the mass spectrum, we need the threshold value as a function of the PBH mass $M$. 
For this purpose, let us consider the horizon entry condition. 
In Eq. (\ref{maxth}), we have implicitly used Eq. (\ref{onehalf}) with the horizon entry condition 
\begin{equation}
aH=\frac{a}{R(\bar r_{\rm m})}=\frac{1}{\bar r_{\rm m}}\ee^{\mu g_{\rm m}}.  \label{hentry}
\end{equation}
We note that this coincides with the condition $2M_{\rm F}(\bar r_{\rm m}\ee^{-\bar\zeta(\bar r_{\rm m})})=H^{-1}$. 
Since the PBH mass is
given by $M=\alpha/(2H)$ with $\alpha$ being a numerical factor, from
the horizon entry condition (\ref{hentry}), the PBH mass $M$ can be expressed as
follows:
\begin{eqnarray}
M&=&\frac{1}{2}\alpha H^{-1}=\frac{1}{2}\alpha a\bar r_{\rm m}\ee^{-\mu g_{\rm m}}
=M_{\rm eq}k_{\rm eq}^2\bar r_{\rm m}^2\ee^{-2\mu g_{\rm m}}=:M^{(\mu,k_*)}(\mu,k_*), 
\label{eq:kM}
\end{eqnarray}
where we have used the fact $H\propto a^{-2}$ and
$a =a_{\rm eq}^2 H_{\rm eq} \bar r_{\rm m}\ee^{-\mu g_{\rm m}}$ 
with $a_{\rm eq}$ and
$H_{\rm eq}$ being the scale factor and Hubble expansion rate at the
matter radiation equality.  
$M_{\rm eq}$ and $k_{\rm eq}$ are defined
by $M_{\rm eq}=\alpha H_{\rm eq}^{-1}/2$ and
$k_{\rm eq}=a_{\rm eq}H_{\rm eq}$, respectively. 
We have also introduced the function $M^{(\mu,k_*)}(\mu,k_*)$. 
The value of the numerical factor $\alpha$ is rather ambiguous, and we 
set $\alpha=1$ as a fiducial value
\footnote{
\baselineskip5mm
If we take into account the critical behavior\cite{Choptuik:1992jv,Koike:1995jm} near the PBH formation threshold, 
$\alpha$ should be given by a function of $\mu$ and $k_*$ as 
$\alpha= K(k_*)(\mu-\mu_{\rm th}(k_*))^\gamma$ with $\gamma\simeq 0.36$~\cite{Niemeyer:1997mt,Yokoyama:1998xd,Niemeyer:1999ak,Green:1999xm,Musco:2004ak,Musco:2008hv,Musco:2012au,Kuhnel:2015vtw,Germani:2018jgr}. 
Since the profile depenence of the function $K(k_*)$ and the parameter range of the scaling behavior 
is not well understood, for simplicity, we just treat $\alpha$ as a numerical factor which takes 
a typical value of order 1 in this paper. 
}. 
%

Then, we may obtain the threshold value of $\mu_{\rm th}^{(M)}(M)$ 
as a function of $M$ by eliminating $k_*$ from Eqs.~\eqref{eq:kM} and $\mu=\mu^{(k_*)}_{\rm th}(k_*)$ and 
solving it for $\mu$. 
That is, defining 
$k_*^{\rm th}(M)$ by the inverse function of 
$M=M^{(\mu,k_*)}(\mu_{\rm th}^{(k_*)}(k_*),k_*)$, 
we obtain the threshold value of $\mu_{\rm th}^{(M)}$ for a fixed value of $M$ as 
\begin{equation}
\mu_{\rm th}^{(M)}(M):=\mu_{\rm th}^{(k_*)}(k_*^{\rm th}(M)). 
\label{eq:muthM}
\end{equation}
While, from Eq.~\eqref{eq:kM}, we can describe $\mu$ as a function of $M$ and $k_*$ as follows:
\begin{equation}
\mu=
\mu^{(M,k_*)}(M,k_*):=
-\frac{1}{2g_{\rm m}}\ln \left(\frac{1}{k_{\rm eq}^2\bar r_{\rm m}^2}\frac{M}{M_{\rm eq}}\right). 
\end{equation}
As is explicitly shown for an extended power spectrum, the value of $\mu$ may be bounded below 
by $\mu_{\rm min}(M)$ for a fixed value of $M$. 
Then, for a fixed value of $M$, the region of $\mu$ for PBH formation can be given by 
\begin{equation}
\mu>\mu_{\rm b}:=\max\left\{\mu_{\rm min}(M),\mu_{\rm th}^{(M)}(M)\right\}. 
\label{eq:mub}
\end{equation}
\section{random Gaussian distribution of $\zeta$}
\label{Gauss}

A key assumption is the random Gaussian distribution of $\zeta$ with 
its power spectrum $\mathcal P(k)$. 
In this section, we briefly review Ref.~\cite{1986ApJ...304...15B} 
to introduce the probability distribution for the curvature variables. 
Due to the random Gaussian assumption, the probability distribution of any set of 
linear combination of the variable $\zeta(x_i)$ is given by a multi-dimensional Gaussian 
probability distribution~\cite{1970Afz.....6..581D,1986ApJ...304...15B}: 
\begin{equation}
P(V_I)\dd^n V_I=\left(2\pi\right)^{-n/2}\left|\det \mathcal M\right|^{-1/2}
\exp \left[-\frac{1}{2}V_I \left(\mathcal M^{-1}\right)^{IJ}V_J\right]\dd^n V, 
\end{equation}
where the components of the matrix $\mathcal M$ are given by 
the correlation $<V_I V_J>$ defined by 
\begin{equation}
<V_I V_J>:=\int \frac{\dd \bm k}{(2\pi)^3}\frac{\dd \bm k'}{(2\pi)^3}<\tilde V_I^*(\bm k)\tilde V_J(\bm k')> 
\end{equation}
with $\tilde V_I(\bm k)=\int \dd^3 xV_I(\bm x)\ee^{i\bm k\bm x}$. 

Here, we consider the value of $\zeta$ and its derivatives up to the second order 
of the Taylor expansion of the field $\zeta(x_i)$:
\begin{equation}
\zeta=\zeta_0+\zeta_1^ix_i+\frac{1}{2}\zeta_2^{ij}x_ix_j
+\mathcal O(x^3). 
\label{eq:taylor}
\end{equation}
The non-zero correlations between two of $\zeta_0$, $\zeta_1^i$ and $\zeta_2^{ii}$ are 
given by 
\begin{eqnarray}
<\zeta_0\zeta_0>&=&
\sigma_0^2, 
\label{eq:sig0}\\
-3<\zeta_0\zeta_2^{ii}>&=&3<\zeta_1^i\zeta_1^i>=
\sigma_1^2, 
\label{eq:sig1}\\
5<\zeta_2^{ii}\zeta_2^{ii}>&=&15<\zeta_2^{ii}\zeta_2^{jj}>=15<\zeta_2^{ij}\zeta_2^{ij}>
=
\sigma_2^2~{\rm with}~i\neq j.  
\label{eq:sig2}
\end{eqnarray}

Let us focus on the variables $\zeta_2^{ij}$.  There are 6 independent
variables $\zeta_2^A:=(\zeta_2^{11}$, $\zeta_2^{22}$, $\zeta_2^{33}$,
$\zeta_2^{12}$, $\zeta_2^{23}$, $\zeta_2^{31})$.  It can be shown
that, taking the principal direction of the matrix $\zeta_2^{ij}$, the
volume element can be rewritten as follows:
\begin{equation}
\prod_{A=1}^6\dd\zeta_2^A=(\lambda_1-\lambda_2)(\lambda_2-\lambda_3)(\lambda_1-\lambda_3)
\sin\theta_1
\dd \lambda_1\dd\lambda_2\dd\lambda_3
\dd \theta_1\dd \theta_2\dd \theta_3, 
\end{equation}
where $\lambda_i$ are eigen values of the matrix $\zeta_2^{ij}$ with $\lambda_1\geq\lambda_2\geq\lambda_3$ 
and $\theta_i$ are the Euler angles to take the principal direction. 
From the integration with respect to the Euler angles, the factor $2\pi^2$ arises.

Following Ref.~\cite{1986ApJ...304...15B}, we introduce 
 new variables $\nu$, $\eta_i$ and $\xi_i$ as follows:
\begin{eqnarray}
\nu&=&-\zeta_0/\sigma_0, \\
\eta_i&=&\zeta_1^i/\sigma_1, \\
\xi_1&=&\left(\lambda_1+\lambda_2+\lambda_3\right)/\sigma_2, \\
\xi_2&=&\frac{1}{2}\left(\lambda_1-\lambda_3\right)/\sigma_2, \\
\xi_3&=&\frac{1}{2}\left(\lambda_1-2\lambda_2+\lambda_3\right)/\sigma_2. 
\end{eqnarray}
$\lambda_i$ is described in terms of $\xi_i$ as follows:
\begin{eqnarray}
\lambda_1&=&\frac{1}{3}\left(\xi_1+3\xi_2+\xi_3\right)\sigma_2, \\
\lambda_2&=&\frac{1}{3}\left(\xi_1-2\xi_3\right)\sigma_2, \\
\lambda_3&=&\frac{1}{3}\left(\xi_1-3\xi_2+\xi_3\right)\sigma_2. 
\end{eqnarray}
Then, the probability distribution can be expressed as
\begin{equation}
P(\nu,\bm \xi,\bm \eta)\dd\nu\dd\bm\xi\dd\bm\eta
= P_1(\nu,\xi_1) P_2(\xi_2,\xi_3,\bm \eta)\dd \nu\dd\bm\xi\dd\bm\eta,
\end{equation}
where
\begin{eqnarray}
&& P_1(\nu,\xi_1)\dd\nu\dd\xi_1=\frac{1}{2\pi}\frac{1}{\sqrt{1-\gamma^2}}
\exp\left[-\frac{1}{2}\left(\nu^2+\frac{(\xi_1-\gamma\nu)^2}{1-\gamma^2}\right)\right]\dd\nu\dd\xi_1
\label{eq:p1}
\end{eqnarray}
and 
\begin{eqnarray}
P_2(\bm \eta, \xi_2,\xi_3)\dd\xi_2\dd\xi_3\dd \bm \eta&=&
\frac{5^{5/2}3^{7/2}}{(2\pi)^2}
\xi_2(\xi_2^2-\xi_3^2)\exp\left[-\frac{5}{2}\left\{
3\xi_2^2+\xi_3^2\right\}\right] 
\cr
&&\times
\exp\left[-\frac{3}{2}\left\{\eta_1^2+\eta_2^2+\eta_3^2\right\}\right]\dd\xi_2\dd\xi_3\dd\bm\eta 
\end{eqnarray}
with 
\begin{equation}
\gamma=\sigma_1^2/(\sigma_0\sigma_2),  
\end{equation}
$\xi_2\geq\xi_3\geq-\xi_2$ and $\xi_2\geq0$. 
We have abbreviated the three components variables $\xi_i$ and $\eta_i$ as $\bm \xi$ and $\bm \eta$. 
In terms of $\mu=\nu\sigma_0$ and $k_*^2=\xi_1\sigma_2/\mu$, the probability $P_1$ can be expressed as 
\begin{eqnarray}
P_1(\nu,\xi_1)\dd\nu\dd\xi_1&=&\frac{\mu}{2\pi\sigma_0\sigma_2\sqrt{1-\gamma^2}}
\exp\left[-\frac{1}{2}\mu^2\left(\frac{1}{\sigma_0^2}+\frac{1}{\sigma_2^2}\frac{(k_*^2-k_{\rm c}^2)^2}{1-\gamma^2}\right)\right]\dd\mu\dd k_*^2,\cr
&=&
\frac{\mu}{2\pi\sigma_0\sigma_2\sqrt{1-\gamma^2}}
\exp\left[-\frac{1}{2}\frac{\mu^2}{\tilde \sigma(k_*)^2}\right]\dd\mu\dd k_*^2,
\label{eq:p1mod}
\end{eqnarray}
where $k_{\rm c}^2=\gamma\sigma_2/\sigma_0=\sigma_1^2/\sigma_0^2$, and
\begin{equation}
\frac{1}{\tilde\sigma(k_*)^2}=\frac{1}{\sigma_0^2}+\frac{\left({k_*}^2-k_{\rm c}^2\right)^2}{(1-\gamma^2)\sigma_2^2}. 
\label{eq:sigtil}
\end{equation} 

\section{PBH fraction to the total density}

\subsection{General expression}
\label{general}

Following Ref.~\cite{1986ApJ...304...15B}, we start by deriving an expression for the peak number density. 
The probability distribution can be written as 
\begin{eqnarray}
&&P (\nu,\bm \eta, \bm \xi)\dd \nu \dd \bm \eta \dd \bm \xi=
P_1(\nu, \xi_1)\dd \nu \dd \xi_1 ~
P_2(\bm \eta, \xi_2, \xi_3)\dd \bm \eta \dd \xi_2 \dd \xi_3. 
\end{eqnarray}
Let us focus on the parameters $\nu$ and $\bm \xi$ 
to characterize each extremum. 
We define $n_{\rm ext}(\bm x,\nu, \xi_1)$ as the distribution of extrema of the field $\zeta$ 
in the space of $(\bm x,\nu, \xi_1)$, that is, 
\begin{equation}
n_{\rm ext}(\bm x,\nu, \xi_1)\Delta \bm x \Delta \nu \Delta \xi_1 ={\rm number~of~extrema~in~the~volume~}\Delta \bm x \Delta \nu \Delta \xi_1. 
\end{equation}
Then, $n_{\rm ext}(\bm x,\nu, \xi_1)$ can be 
expressed as follows:
\begin{equation}
n_{\rm ext}(\bm x,\nu, \xi_1)\dd \bm x \dd\nu \dd \xi_1 
=\sum_p\delta(\bm x-\bm x_p)\delta(\nu-\nu_p)\delta(\xi_1-\xi_{1p})\dd \bm x \dd \nu 
\dd \xi_1, 
\end{equation}
where we have expressed the variables at each extremum with the subscript $p$. 
Then, $\bm x_p$ is the position of the extremum, that is, $\bm \zeta_1=\bm \eta=0$ at $\bm x=\bm x_p$. 
Therefore, we obtain 
\begin{equation}
\delta(\bm x-\bm x_p)=\det \left|\frac{\del^2 \zeta}{\del x^i\del x^j}\right|_{\bm x=\bm x_p}\delta(\bm \zeta_1)=
\sigma_1^{-3}|\lambda_1\lambda_2\lambda_3|\delta(\bm \eta), 
\end{equation}
where 
\begin{equation}
\lambda_1\lambda_2\lambda_3=\frac{1}{27}\left(\left(\xi_1+\xi_3\right)^2-9\xi_2^2\right)(\xi_1-2\xi_3)\sigma_2^3. 
\label{eq:xeta}
\end{equation}
The peak number density $n_{\rm pk}(\nu,\xi_1)$ is given by the ensemble average of 
$n_{\rm ext}\Theta(\lambda_3)$ 
as follows:\footnote{\baselineskip5mm
\dred{We are very grateful to Jianing Wang, Shi Pi, Misao Sasaki and 
Volodymyr Takhistov for pointing out the error in the coefficient of Eq.~\eqref{eq:corrected_npk}
in the previous version of this paper. }}
\begin{eqnarray}
n_{\rm pk}(\nu,\xi_1)\dd \nu\dd\xi_1&=&<n_{\rm ext}\Theta(\lambda_3)>\dd \nu\dd\xi_1
\cr
&=&\sigma_1^{-3}\Biggl[\int \dd \nu_p \dd \xi_{1p} \dd \bm \eta \dd\bm\xi 
\Bigl\{P(\nu_p,\bm\eta,\xi_{1p},\xi_2,\xi_3) |\lambda_1\lambda_2\lambda_3|
\cr
&&\hspace{3cm}\delta(\bm \eta)\delta(\nu-\nu_p)\delta(\xi_1-\xi_{1p})\Theta(\lambda_3)
\Bigr\}\Biggr]
\dd \nu\dd\xi_1
\cr
&=&
\dred{3^{-3/2}(2\pi)^{-3/2}}
\left(\frac{\sigma_2}{\sigma_1}\right)^3
f(\xi_1)P_1(\nu,\xi_1)\dd\nu \dd\xi_1, 
\label{eq:corrected_npk}
\end{eqnarray}
where $\Theta(\lambda_3)$ is multiplied to pick peaks out of extrema, and 
the function $f$ is given by 
\begin{eqnarray}
f(\xi_1)&:=&\dred{\frac{3^25^{5/2}}{\sqrt{2\pi}}}
\left(\int^{\xi_1/4}_0\dd \xi_2\int^{\xi_2}_{-\xi_2}\dd \xi_3
+\int^{\xi_1/2}_{\xi_1/4}\dd \xi_2\int^{\xi_2}_{3\xi_2-\xi_1}\dd \xi_3
\right)\cr
&&\left\{\xi_2(\xi_2^2-\xi_3^2)\left\{(\xi_1+\xi_3)^2-9\xi_2^2\right\}(\xi_1-2\xi_3)
\exp\left[-\frac{15}{2}\xi_2^2\right]
\exp\left[-\frac{5}{2}\xi_3^2\right]\right\}\cr
&=&\frac{1}{2}\xi_1(\xi_1^2-3)
\left({\rm erf}\left[\frac{1}{2}\sqrt{\frac{5}{2}}\xi_1\right]
+{\rm erf}\left[\sqrt{\frac{5}{2}}\xi_1\right]\right)\cr
&&\hspace{1cm}+\sqrt{\frac{2}{5\pi}}\left\{
\left(\frac{8}{5}+\frac{31}{4}\xi_1^2\right)
\exp\left[-\frac{5}{8}\xi_1^2\right]
+\left(-\frac{8}{5}+\frac{1}{2}\xi_1^2\right)\exp\left[-\frac{5}{2}\xi_1^2\right]
\right\}. 
\label{eq:funcf}
\end{eqnarray}

We note that, due to the condition $\lambda_3>0$, we obtain
$\zeta_2=\xi_1\sigma_2>0$.  
Let us change the 
variables from $\nu=-\zeta_0/\sigma_0=\mu/\sigma_0$ and 
$\xi_1=\triangle \zeta|_{\rm r=0}/\sigma_2=\mu k_*^2/\sigma_2$ to 
variables $\mu$ and $k_*$.  
Then, we obtain 
\begin{eqnarray}
&&n^{(k_*)}_{\rm pk}(\mu, k_*) \dd \mu\dd k_*
:=n_{\rm pk}(\nu,\xi_1)\dd \nu\dd\xi_1\cr
&&\hspace{2cm}=\dred{2 \cdot 3^{-3/2}(2\pi)^{-3/2}}\mu k_*\frac{\sigma_2^2}{\sigma_0\sigma_1^3}f\left(\frac{\mu k_*^2}{\sigma_2}\right)
P_1\left(\frac{\mu}{\sigma_0},\frac{\mu k_*^2}{\sigma_2}\right) 
\dd\mu\dd k_*.  
\label{eq:nks}
\end{eqnarray}

Since the direct observable is not $k_*$ but the PBH mass $M$, we
further change the variable from $k_*$ to $M$ as follows:
\begin{eqnarray}
&&n^{(M)}_{\rm pk}(\mu, M)\dd \mu\dd M
:=n^{(k_*)}_{\rm pk}(\mu, k_*)\dd \mu\dd k_*\cr
&&=\dred{3^{-3/2}(2\pi)^{-3/2}}
\frac{\sigma_2^2}{\sigma_0\sigma_1^3}
\mu k_*
f\left(\frac{\mu k_*^2}{\sigma_2}\right)
P_1\left(\frac{\mu}{\sigma_0},\frac{\mu k_*^2}{\sigma_2}\right) 
\left|\frac{\dd}{\dd k_*}\ln \bar r_{\rm m}-\mu \frac{\dd}{\dd k_*}g_{\rm m}\right|^{-1}
\dd\mu\dd \ln M,  
\end{eqnarray}
where $k_*$ should be regarded as a function of $\mu$ and $M$ 
given by solving Eq.~\eqref{eq:kM} for $k_*$. 
Here, we note that an extended power spectrum is implicitly assumed in the above expression. 
The monochromatic spectrum case will be independently discussed in Sec.~\ref{sec:monochrom}. 
We also note that, since we relate $k_*$ to $M$ with $\mu$ fixed, 
we have implicitly assumed that there is only one peak with $\triangle \zeta=\mu k_*^2$ 
in the region corresponding to the mass $M$, that is, inside $r=r_{\rm m}$. 
If the spectrum is broad enough or has multiple peaks at far separated scales, 
and the typical PBH mass is relatively larger than the minimum scale given by the spectrum, 
we would find multiple peaks inside $r=r_{\rm m}$. 
Then, the PBH abundance would be overestimated because we count every peak as a candidate for 
a PBH formation. 
In order to avoid this difficulty, 
we simply assume that the power spectrum is characterized by a single scale $k_0$ and 
has a localized peak around the scale $k_0$. 
Therefore, the current version of our procedure cannot be directly applied to a spectrum with a broad support or 
multiple scales.

The number density of PBHs is given by 
\begin{equation}
n_{\rm BH} \dd \ln M=\left[
\int^\infty_{\mu_{\rm b}}\dd \mu
~n^{(M)}_{\rm pk}(\mu,M)\right]M \dd \ln M. 
\end{equation}
We also note that the scale factor $a$ is a function of $M$ as $a=2M^{1/2}M_{\rm eq}^{1/2}k_{\rm eq}/\alpha$. 
Then, the fraction of PBHs to the total density $\beta_0\dd \ln M$ can be given by 
\begin{eqnarray}
\beta_0\dd \ln M&=&\frac{M n_{\rm BH}}{\rho a^3}\dd \ln M
=\frac{4\pi}{3}\alpha n_{\rm BH}k_{\rm eq}^{-3}\left(\frac{M}{M_{\rm eq}}\right)^{3/2} \dd \ln M\cr
&=&\dred{\frac{2\alpha k_{\rm eq}^{-3}}{3^{5/2}(2\pi)^{1/2}}}
\frac{\sigma_2^2}{\sigma_0\sigma_1^3}
\left(\frac{M}{M_{\rm eq}}\right)^{3/2}
\Biggl[
\int^\infty_{\mu_{\rm b}}\dd\mu 
\mu k_*f\left(\frac{\mu k_*^2}{\sigma_2}\right)
\cr
&&\hspace{2cm}P_1\left(\frac{\mu}{\sigma_0},\frac{\mu k_*^2}{\sigma_2}\right)
\left|\frac{\dd}{\dd k_*}\ln \bar r_{\rm m}-\mu \frac{\dd}{\dd k_*}g_{\rm m}\right|^{-1}
\Biggr]\dd \ln M. 
\label{eq:beta_general}
\end{eqnarray}
Here we note again that $k_*$ should be regarded as a function of $\mu$ and $M$. 
The above formula can be evaluated in principle once the form of the power spectrum is given. 
The PBH fraction to the total density $f_0$ at the equality time is given by $f_0=\beta_0 (M_{\rm eq}/M)^{1/2}$.

Let us summarize how to use the above formula. 
Once the power spectrum characterized by a single scale $k_0$ is given, 
the typical profile is given by Eq.~\eqref{eq:zetahat}. 
Taking the radius of the maximum compaction function \eqref{eq:Compaction} for this profile, 
we obtain the function $\bar r_{\rm m}(k_*)$ and $g_{\rm m}(k_*)=g(\bar r_{\rm m}(k_*);k_*)$. 
Since $k_*$ is implicitly given as a function of $\mu$ and $M$ as Eq.~\eqref{eq:kM}, 
we can express the integrand in Eq.~\eqref{eq:beta_general} as a function of $\mu$ and $M$.  
The lower boundary $\mu_{\rm b}$ of the integration is given by Eq.~\eqref{eq:mub}, 
where $\mu_{\rm th}^{(M)}(M)$ is implicitly given by eliminating $k_*$ from 
Eqs.~\eqref{eq:muth} and \eqref{eq:kM} as is defined in Eq.~\eqref{eq:muthM}.

From the functional form \eqref{eq:p1mod} of $P_1$, we may expect that the 
integrand of Eq.~\eqref{eq:beta_general} has a non-negligible value only around 
$k=k_{\rm c}$ and $\mu=\mu_{\rm b}$. 
Assuming $\mu_{\rm b}\gg\tilde \sigma$, 
we can obtain the following approximate form without performing the integral:
\begin{eqnarray}
\beta_0^{\rm approx}\dd \ln M
&=&\dred{\frac{2\alpha k_{\rm eq}^{-3}}{3^{5/2}(2\pi)^{1/2}}}
\frac{\sigma_2^2}{\sigma_0\sigma_1^3}
\left(\frac{M}{M_{\rm eq}}\right)^{3/2}
\Biggl[
\tilde \sigma(k_*)^2k_*f\left(\frac{\mu k_*^2}{\sigma_2}\right)
\cr
&&\hspace{2cm}P_1\left(\frac{\mu}{\sigma_0},\frac{\mu k_*^2}{\sigma_2}\right)
\left|\frac{\dd}{\dd k_*}\ln \bar r_{\rm m}-\mu \frac{\dd}{\dd k_*}g_{\rm m}\right|^{-1}
\Biggr]_{\mu=\mu_{\rm b}}\dd \ln M. 
\label{eq:beta_app}
\end{eqnarray}
Again, $k_*$ should be regarded as a function of $\mu$ and $M$ in the above expression.
Let us roughly estimate the typical width $\Delta \ln M$ in the mass spectrum from 
the above approximate expression. 
From the horizon entry condition and a rough dimensional analysis, 
we may estimate the relation between $M$ and $k_*$ as $M\sim H^{-1}\propto k_*^{-2}$. 
Then, we find $\Delta \ln M\sim\Delta \ln k_*^2\lesssim\sigma_2/k_{\rm c}^2\sim \sigma_0$.

The probability $P_1$ takes the maximum value at $k_*=k_{\rm c}$ for a fixed value of 
$\mu$, and the threshold value of $\mu$ for $k_*=k_{\rm c}$ 
is given by $\mu_{\rm c}:=\mu^{(k_*)}_{\rm th}(k_{\rm c})$. 
Then, the value of the mass at the peak probability is given by 
\begin{equation}
M=M_{\rm c}:=M_{\rm eq}k_{\rm eq}^2\bar r_{\rm m}(k_{\rm c})^2\ee^{-2\mu_{\rm c}g_{\rm c}}, 
\label{eq:Mc}
\end{equation}
where $g_{\rm c}:=g_{\rm m}(k_{\rm c})$. 
The value of $\beta_0$ can be evaluated as
\begin{eqnarray}
\beta_0^{\rm approx}(M_{\rm c})
&\simeq&\dred{\frac{2 \alpha \ell^3 \mu_{\rm c}^3 \ee^{-3\mu_{\rm c}g_{\rm c}}}{3^{5/2} (2\pi)^{3/2}}}
\frac{\sigma_1^2}{\sigma_0^4\sigma_2}\exp\left[-\frac{1}{2}\frac{\mu_{\rm c}^2}{\sigma_0^2}\right]
\label{eq:ext_approx}
\\
&\sim&
\frac{\mu_{\rm c}^3}{\sigma_0^3}\ee^{-3\mu_{\rm c}g_{\rm c}}\exp\left[-\frac{1}{2}\frac{\mu_{\rm c}^2}{\sigma_0^2}\right], 
\label{eq:ext_approx_order}
\end{eqnarray}
where $\ell:=\bar r_{\rm m}(k_{\rm c})k_{\rm c} $, and 
we have evaluated as 
$|\frac{\dd}{\dd k_*}\ln \bar r_{\rm m}-\mu \frac{\dd}{\dd k_*}g_{\rm m}|\simeq1/k_*$, 
$f(\mu k_*^2/\sigma_2)\simeq \mu^3 k_*^6/ \sigma_2^3$\cite{1986ApJ...304...15B} in the first line. 
A numerical factor of the order of unity is neglected in the second line, 
and the values of moments are assumed to be $\sigma_0\sim \sigma_1/k_{\rm c}\sim \sigma_2/k_{\rm c}^2$.

Let us summarize how to use the above approximate formula. 
Once the power spectrum is given, the typical profile with $k_*=k_{\rm c}=\sigma_1/\sigma_0$ is given by 
its two point correlation function $\psi(r)$. 
Calculating the compaction function and taking the radius of the maximum $\mathcal C$ for the typical profile, 
we can obtain the value of $\ell$. 
The threshold value $\mu_{\rm c}$ can be evaluated by the formula \eqref{eq:muth} with 
$k_*$ being $k_{\rm c}$, where the value of $\delta_{\rm th}$ should be provided. 
Then, the simplified version of the PBH fraction \eqref{eq:ext_approx} at the spectrum peak 
can be calculated. Therefore, necessary ingredients are the power spectrum and the values 
of $\delta_{\rm th}$ and $\alpha$. 


\subsection{Estimation from the Press-Schechter formalism}
\label{sec:PS}

For a comparison, we review a conventional estimate 
of the fraction of PBHs based on the PS formalism. 
In the conventional formalism, the scale dependence is introduced by a window function $W(k/k_M)$, 
where 
\begin{equation}
k_M =k_{\rm eq}(M_{\rm eq}/M)^{1/2}. \label{eq:kmM}
\end{equation} 
Then, each gradient moment is replaced by the following expression: 
\begin{equation}
\hat\sigma_n(k_*)^2
=\int\frac{\dd k}{k} k^{2n} \mathcal  P(k) W(k/k_M)^2.  
\label{eq:moments_window}
\end{equation}
The conventional estimate starts 
from the following Gaussian distribution assumption for the density perturbation $\bar \delta$: 
\begin{equation}
P_\delta(\bar \delta)\dd \bar\delta=
\frac{1}{\sqrt{2\pi }\sigma_\delta}\exp\left(-\frac{1}{2}\frac{\bar \delta^2}{\sigma_\delta^2}\right)\dd \bar \delta, 
\label{eq:Gaussian_delta}
\end{equation}
where $\sigma_\delta$ is given by the coarse-grained density contrast 
\begin{equation}
\sigma_\delta(k_M)=\frac{4}{9}\frac{\hat \sigma_2(k_M)}{k_M^2}. 
\label{eq:sigdel}
\end{equation}
We note that the definition of $\bar \delta$ and $\delta_{\rm th}$ used in this conventional estimate 
are rather vague and not necessarily identical to our definition of $\bar\delta$ and $\delta_{\rm th}$.
Therefore, there is an ambiguity of which definition of the density perturbation is supposed in this formalism. 
Here, for simplicity, we just use the same numerical value of $\delta_{\rm th}$ as in our approach, 
in other words, we assume that the volume average of the density perturbation 
obeys the Gaussian probability distribution given by Eq.~\eqref{eq:Gaussian_delta} with 
the coarse-grained density contrast \eqref{eq:sigdel} in the PS formalism. 
This Gaussian distribution and the dispersion are motivated by the linear relation between $\zeta$ and $\delta$.
The fraction $\beta_0^{\rm PS}$ is then evaluated as follows(see e.g. \cite{Carr:2016drx}):
\begin{equation}
\beta_{0,\delta}^{\rm PS}(M)=2\alpha \int^\infty_{\delta_{\rm th}}\dd\delta P_\delta(\delta) 
=\alpha
{\rm erfc}\left(\frac{\delta_{\rm th}}{\sqrt{2}\ \sigma_\delta(k_M)}\right)
=\alpha
{\rm erfc}\left(\frac{9}{4}\frac{\delta_{\rm th}k_M^2}{\sqrt{2}\hat \sigma_2(k_M)}\right).
\label{eq:conv0}
\end{equation} 
The PBH fraction to the total density $f_{0,\delta}^{\rm PS}$ at the equality time is
 given by $f_{0,\delta}^{\rm PS}=\beta_{0,\delta}^{\rm PS} (M_{\rm eq}/M)^{1/2}$.

Let us consider another way of estimation as a reference. 
First, referring to the linear relation $\delta=4\triangle \zeta/(9k_*^2)=4\mu/9$ 
at the conventional horizon entry $aH=k_*$, 
we change the variable for the Gaussian distribution Eq.~\eqref{eq:Gaussian_delta} to $\mu$ as 
\begin{equation}
P_\mu(\mu)\dd \mu=
\frac{1}{\sqrt{2\pi }\hat\sigma_2}\exp\left(-\frac{1}{2}\frac{\mu^2k_M^4}{\hat\sigma_2^2}\right)\dd \mu.  
\end{equation}
As in the case of $\beta_{0,\delta}^{\rm PS}$, we can evaluate the PBH fraction $\beta_{0,\mu}^{\rm PS}$
as follows:
\begin{equation}
\beta_{0,\mu}^{\rm PS}(M)=2\alpha \int^\infty_{\mu_{\rm c}}\dd\mu P_\mu(\mu) 
=\alpha
{\rm erfc}\left(\frac{\mu_{\rm c}k_M^2}{\sqrt{2}\ \hat \sigma_2(M)}\right), \label{psmu}
\end{equation} 
where the threshold value $\mu_{\rm c}=\mu^{(k_*)}_{\rm th}(k_{\rm c})$ is evaluated by the procedure 
introduced in Sec.~\ref{sec2}, where the non-linearity is taken into account. 
Therefore, comparing Eq.~\eqref{eq:conv0} and Eq.~\eqref{psmu}, 
we can extract the effect of the optimized criterion proposed in Sec.~\ref{sec2}. 
The PBH fraction to the total density $f_{0,\mu}^{\rm PS}$ at the equality time is
 given by $f_{0,\mu}^{\rm PS}=\beta_{0,\mu}^{\rm PS} (M_{\rm eq}/M)^{1/2}$. 
As will be shown later, we obtain $\mu_{\rm c}\sim0.52$ for the monochromatic spectrum and 
$\sim 0.75$ for a specific model of extended spectra. 
Therefore, the value of $\mu_{\rm c}$ is typically smaller than $9\delta_{\rm th}/4\approx1.20$.
This simple analysis clearly indicates that 
the optimized criterion given in Sec.~\ref{sec2} will significantly increase the PBH fraction compared to the 
conventional estimate \eqref{eq:conv0}.

\section{Specific examples }
\label{sec:examples}

\subsection{Monochromatic power spectrum}
\label{sec:monochrom}
\begin{figure}[htbp]
\begin{center}
\includegraphics[scale=0.63]{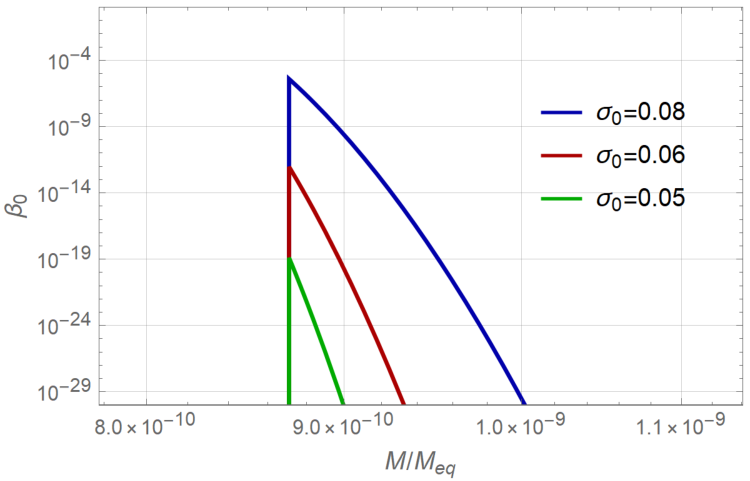}
\includegraphics[scale=0.63]{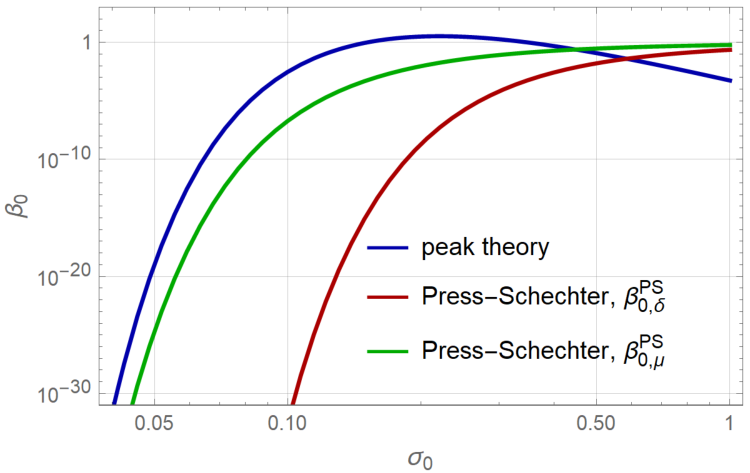}
\caption{
\baselineskip5.5mm
PBH fraction to the total density $\beta_0$ at the formation time 
is depicted as a function of the PBH mass $M$ (left panel) for 
each value of $\sigma_0$, and a function of $\sigma_0$ (right panel) for $M=M_{\rm c}$ with $k_0=10^5k_{\rm eq}$ and $\alpha=1$. 
We also depict the conventional estimation from the PS formalism  without a window function
in the right panel.  
}
\label{fig:beta_mono}
\end{center}
\end{figure}

\begin{figure}[htbp]
\begin{center}
\includegraphics[scale=0.63]{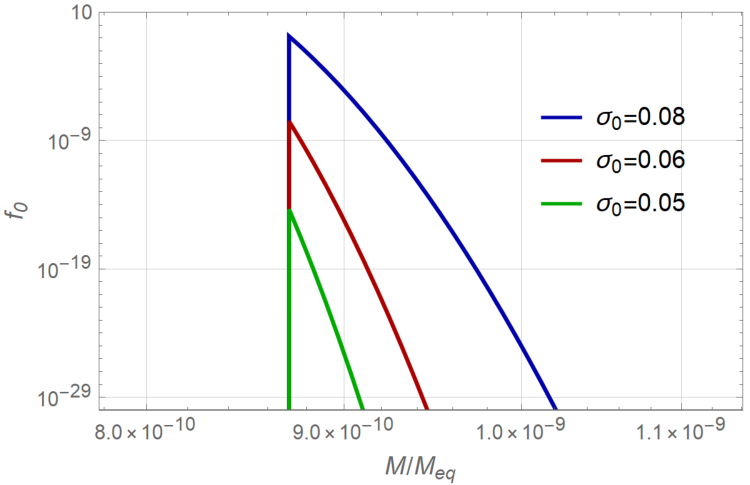}
\includegraphics[scale=0.63]{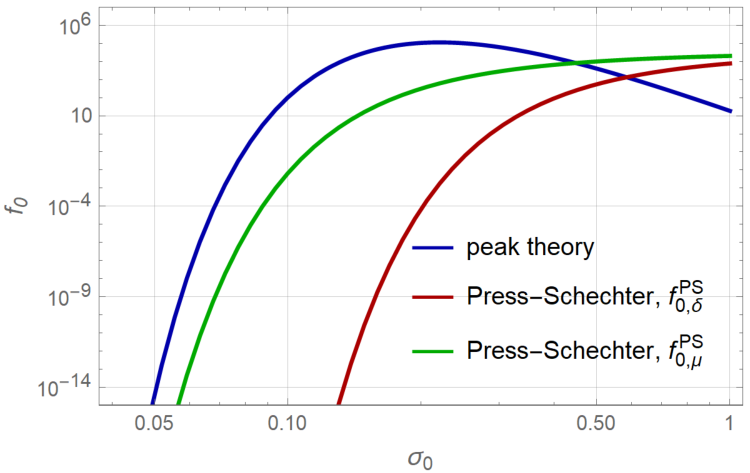}
\caption{
\baselineskip5.5mm
PBH Fraction   to the total density $f_0$ at the equality time 
is depicted as a function of the PBH mass $M$ (left panel) for 
each value of $\sigma_0$,  and  a function of $\sigma_0$ (right panel) for $M=M_{\rm }$ 
with $k_0=10^5k_{\rm eq}$ and $\alpha=1$.  
We also depict the conventional estimation from the PS formalism without a window function 
in the right panel.  
}
\label{fig:f_mono}
\end{center}
\end{figure}

Let us consider the monochromatic power spectrum given by 
\begin{equation}
\mathcal P(k)=\sigma_0^2k_0\delta(k-k_0). 
\end{equation}
Then, the moments are calculated as
\begin{eqnarray}
\sigma_n^2=\sigma^2_0k_0^{2n}. 
\end{eqnarray}
It leads to $k_{\rm c}=k_0$ and $\gamma=1$. 
Since $k_{\rm c}=k_0$, from Eq.~\eqref{eq:gkc}, 
the functional form of $g(r;k_0)$ is given by 
\begin{equation}
g(r;k_0)=-\psi(r)=-\frac{\sin(k_0r)}{k_0r}. 
\end{equation}
Then, we can find $\bar r_{\rm m}(k_0)=\ell/k_0=2.74/k_0$, $\mu_{\rm c}=0.520$ and $g_{\rm c}=-0.141$. 
Since the value of $\gamma$ is given by 1. 
Taking the limit $\gamma\rightarrow1$, in the expression \eqref{eq:p1}, 
we obtain 
\begin{eqnarray}
\lim_{\gamma\rightarrow1}P_1(\nu,\xi_1)
&=&\frac{1}{\sqrt{2\pi}}\delta(\xi_1-\gamma\nu)\exp\left(-\frac{1}{2}\nu^2\right)\cr
&=&\frac{1}{2\sqrt{2\pi}}\frac{\sigma_2}{\mu k_*}\delta(k_*-k_0)\exp\left(-\frac{\mu^2}{2\sigma_0^2}\right). 
\end{eqnarray}
Then, the $k_*$ integration in Eq.~\eqref{eq:nks} can be performed, and we obtain the following 
expression for the peak number density:
\begin{equation}
n^{(\mu)}_{\rm pk}(\mu)\dd \mu=\dred{3^{-3/2}(2\pi)^{-2}}\frac{1}{\sigma_0}k_0^3f\left(\frac{\mu}{\sigma_0}\right)
\exp\left(-\frac{\mu^2}{2\sigma_0^2}\right)\dd \mu. 
\end{equation}
Under the condition $k_*=k_0$, the relation between $M$ and $\mu$ is given by 
\begin{equation}
\mu=\mu_0(M):=-\frac{1}{2g_{\rm c}}\ln\left(\frac{M}{M_{\rm eq}}\frac{k_0^2}{k_{\rm eq}^2}\frac{1}{\ell^2}\right). 
\end{equation}
Therefore, we obtain the following PBH number density:
\begin{equation}
n_{\rm BH}\dd \ln M
=\dred{\frac{k_0^3}{2\cdot 3^{3/2}(2\pi)^{2}\sigma_0|g_{\rm c}|}}f\left(\frac{\mu_0(M)}{\sigma_0}\right)
\exp\left(-\frac{\mu_0(M)^2}{2\sigma_0^2}\right)
\Theta\left(M-M_{\rm c}\right)\dd \ln M, 
\end{equation}
where $\Theta\left(M-M_{\rm c}\right)$ has been multiplied to extract the distribution above the threshold. 
Finally, 
we obtain 
\begin{eqnarray}
\beta_0\dd \ln M&=&
\dred{\frac{\alpha}{3^{5/2}2\pi\sigma_0|g_{\rm c}|}}
\left(\frac{M}{M_{\rm eq}}\right)^{3/2}
\left(\frac{k_0}{k_{\rm eq}}\right)^3
f\left(\frac{\mu_0(M)}{\sigma_0}\right)\cr
&&\hspace{1cm}\exp\left(-\frac{\mu_0(M)^2}{2\sigma_0^2}\right)\Theta\left(M-M_{\rm c}\right)\dd\ln M.  
\end{eqnarray}
At $M=M_{\rm c}=M_{\rm eq}k_{\rm eq}^2\ell^2 k_0^{-2}\ee^{-2\mu_{\rm c}g_{\rm c}}$, the value of $\beta_0$ is given by 
\begin{equation}
\left.\beta_0\right|_{M=M_{\rm c}+0}=\dred{\frac{\alpha}{3^{5/2}2\pi \sigma_0}}\frac{\ell^3}{|g_{\rm c}|}
\ee^{-3\mu_{\rm c}g_{\rm c}}f\left(\frac{\mu_{\rm c}}{\sigma_0}\right)\exp\left(-\frac{\mu_{\rm c}^2}{2\sigma_0^2}\right). 
\end{equation}
Since the function $f(x)$ behaves like $f(x)\sim x^3$ for $x\gg1$~\cite{1986ApJ...304...15B}, 
in the limit of small $\sigma_0$, we obtain 
\begin{equation}
\beta_0\sim \frac{\mu_{\rm c}^3}{\sigma_0^4}\ee^{-3\mu_{\rm c}g_{\rm c}} \exp\left(-\frac{\mu_{\rm c}^2}{2\sigma_0^2}\right), 
\label{eq:beta0monoapprox}
\end{equation}
where a numerical factor of the order of unity is neglected. 
The mass spectra $\beta_0$ and $f_0$ are depicted as functions of the PBH mass $M$ 
for $\sigma_0=0.08$, $0.06$ and $0.05$ in the left panels of Figs.~\ref{fig:beta_mono} and \ref{fig:f_mono}, 
while $\beta_0$ and $f_0=\left(M_{\rm eq}/M\right)^{1/2}\beta_0$ 
 are depicted as functions of $\sigma_0$ for $M=M_{\rm c}$ 
in the right panels of Figs.~\ref{fig:beta_mono}
and \ref{fig:f_mono}. 
%
%
As a result, we obtain an extended mass spectrum of PBHs even for the monochromatic power spectrum of 
the curvature perturbation. 
This is caused by the non-linear relation between the density perturbation and the curvature perturbation 
through Eq.~\eqref{eq:kM}
\footnote{
\baselineskip5mm
The scaling behavior of the PBH mass near the threshold also induces a broadening of the mass spectrum 
as is discussed in Refs.~\cite{Niemeyer:1997mt,Yokoyama:1998xd,Green:1999xm,Kuhnel:2015vtw,Germani:2018jgr}. 
}. 
However, the width of the spectrum is hardly significant. 
It is worthy to be mentioned that comparing Eq.~\eqref{eq:ext_approx_order} with Eq.~\eqref{eq:beta0monoapprox}, 
we find the factor $1/\sigma_0$ of difference. 
This clearly shows that, the monochromatic spectrum case gets larger peak amplitude in the mass spectrum instead of 
the loss of the mass spectrum width.

Note that the amplitude of the mass spectrum is huge compared to
the conventional one $\beta_{0,\delta}^{\rm PS}$ for a small value of $\sigma_0$. 
The main reason for this deviation comes from the 
optimization of the PBH formation threshold $\mu_{\rm c}$. 
The $\sigma_0^{-4}$ dependence of the prefactor in Eq.~\eqref{eq:beta0monoapprox} also 
contributes to increase the fraction, but not so dramatically. 
We note that the strong enhancement in the abundance of PBH is a robust feature, for any chosen value of the threshold 
$\delta_{\rm th}$. Indeed, it follows from (\ref{eq:muth}) that for $\delta_{\rm th}<\delta_{\rm max}=2/3$, we have
$\mu_{\rm c} < (9/4) \delta_{\rm th}$. According to Eqs. (\ref{eq:conv0}) and (\ref{psmu}), this implies 
$\beta_{0,\mu}^{\rm PS}(M) \gg \beta_{0,\delta}^{\rm PS}(M)$. In fact,
for $\delta_{\rm th} < 0.623$, we have $\mu_{\rm c} < (9/8) \delta_{\rm th}$ (the right hand side of this inequality is actually a good 
linear fit to the value of $\mu_{\rm c}$, which typically overestimates the actual value by less than 30\%). Assuming that 
the probability of PBH formation is low, we can approximate the complementary error function as ${\rm erfc}(x)\approx \ee^{-x^2}/(\sqrt{\pi}x)$, and 
it follows that within this range of $\delta_{\rm th}$ we have 
\begin{equation}
\beta_{0,\mu}^{\rm PS}(M) > 
\left(\beta_{0,\delta}^{\rm PS}(M)\right)^{1/4}
\gg \beta_{0,\delta}^{\rm PS}(M),
\end{equation}
where we have ignored the sub-exponential dependence. Given that the probability of PBH formation is exponentially small, this represents a very strong enhancement.

Furthermore, there is a non-trivial correction to the expression for the mass in terms of the wave number 
at the time of horizon crossing, which increases the mass of the black holes by one order of 
magnitude or so relative to the expectation from linear theory. This is clear from the 
expression of $M_{\rm c}$ given in Eq. (\ref{eq:Mc}), which contains 
the factor $\ell^2\ee^{-2\mu_{\rm c}g_{\rm c}}\approx 8.7$. Here, we have used the numerical values for 
$\mu_{\rm c}$, $\ell$ and $g_{\rm c}$ corresponding to the $\rm sinc$ profile, which is the appropriate one for the monochromatic spectrum. 
These values are listed in Table I of Appendix C. Hence, not only do we find more PBHs, but they are 
also significantly larger than naively expected.

\subsection{An extended power spectrum}
\label{sec:extended}

\begin{figure}[htbp]
\begin{center}
\includegraphics[scale=0.63]{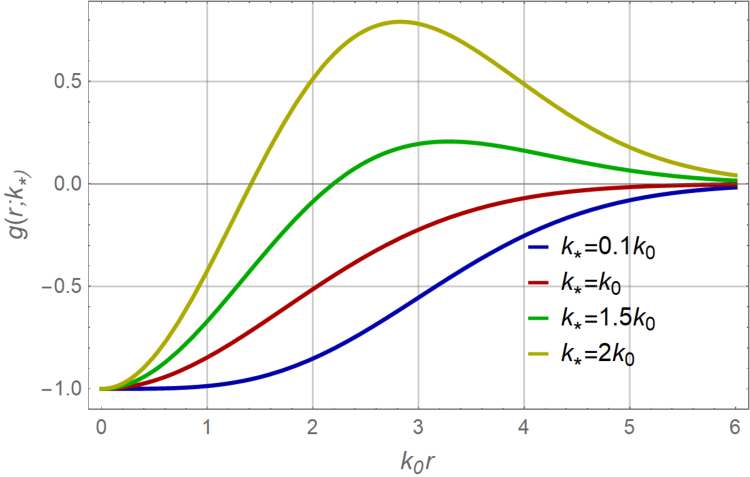}
\includegraphics[scale=0.63]{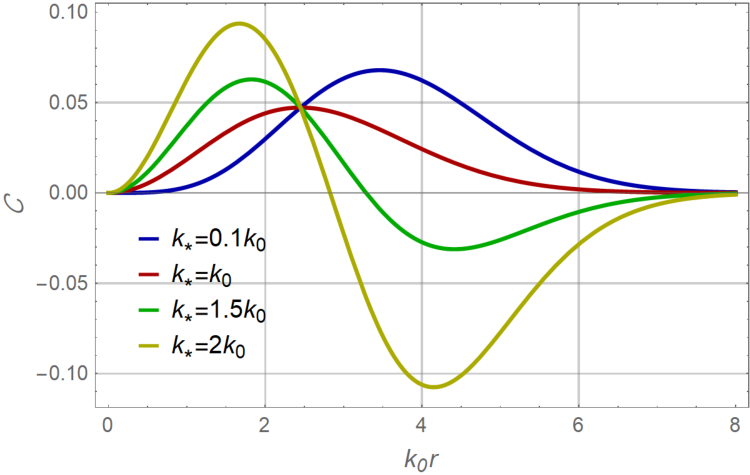}
\caption{$g(r;k_*)$ and $\mathcal C(r)$ for $k_*=0.1k_0$, $k_0$, $1.5k_0$ and $2k_0$. 
The value of $\mu$ is set as $\mu=0.1$ for the compaction function. 
}
\label{fig:hatzeta}
\end{center}
\end{figure}

\begin{figure}[htbp]
\begin{center}
\includegraphics[scale=0.63]{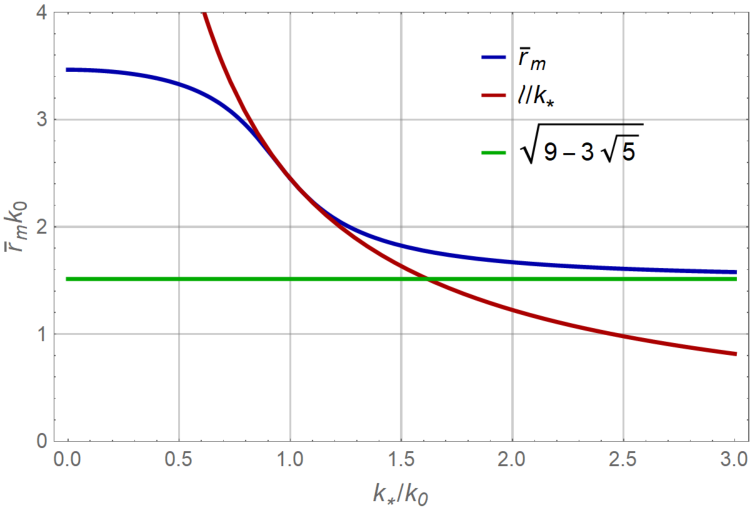}
\caption{$\bar r_{\rm m}$ and the radius of the inner peak of $\mathcal C(r)$. 
}
\label{fig:rm}
\end{center}
\end{figure}

\begin{figure}[htbp]
\begin{center}
\includegraphics[scale=0.63]{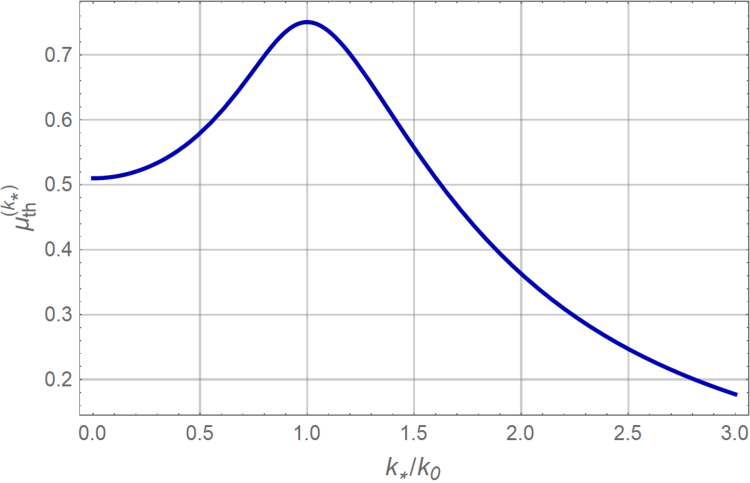}
\includegraphics[scale=0.63]{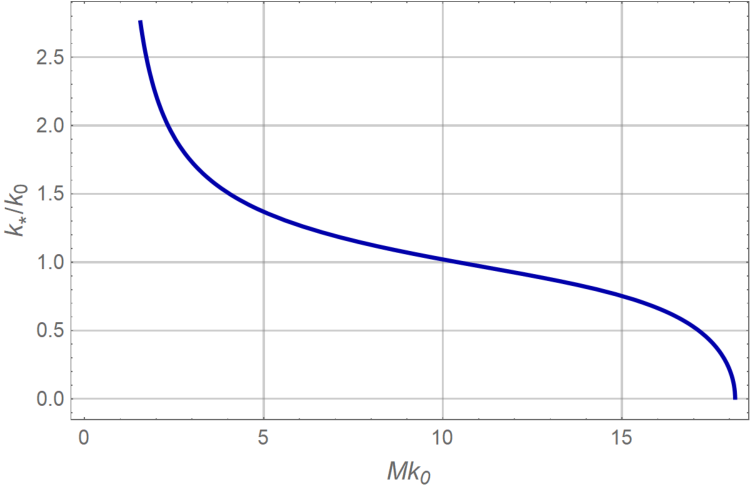}
\caption{$\mu^{(k_*)}_{\rm th}$ as a function of $k_*$(left) and 
$k_*^{\rm th}(M)$ as a function of $M$(right). 
}
\label{fig:muthk}
\end{center}
\end{figure}

\begin{figure}[htbp]
\begin{center}
\includegraphics[scale=0.79]{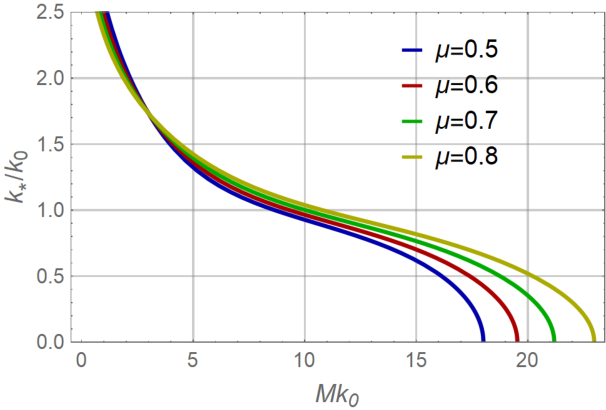}
\caption{The line of $M=M^{(\mu,k_*)}(\mu,k_*)$ for each value of $\mu$. 
}
\label{fig:kM}
\end{center}
\end{figure}

\begin{figure}[htbp]
\begin{center}
\includegraphics[scale=0.79]{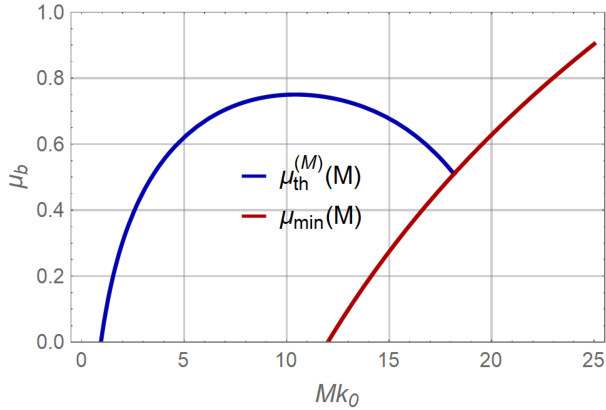}
\caption{
$\mu^{(M)}_{\rm th}$ as a function of $M$. 
}
\label{fig:muthM}
\end{center}
\end{figure}

Let us consider the simple extended power spectrum given by 
\begin{equation}
\mathcal  P(k)=3\sqrt{\frac{6}{\pi}}\sigma^2\left(\frac{k}{k_0}\right)^3
\exp\left(-\frac{3}{2}\frac{k^2}{k_0^2}\right). 
\label{eq:Gausspower}
\end{equation}
Gradient moments are calculated as 
\begin{equation}
\sigma_n^2=\frac{2^{n+1}}{3^n\sqrt{\pi}}\Gamma\left(\frac{3}{2}+n\right)\sigma^2k_0^{2n}. 
\end{equation}
where $\Gamma$ means the gamma function%
\footnote{The overall factor and the power of the exponential is chosen so that we may have 
$\sigma_0=\sigma$ and $k_{\rm c}=k_0$.}. 
The functional form of $\psi(r)$ is given by 
\begin{equation}
\psi(r)=\exp\left(-\frac{k_0^2r^2}{6}\right). 
\end{equation}
Functional forms of $g(r;k_*)$ and the corresponding $\mathcal C(r)$
obtained by substituting $\zeta(r)=\bar{\zeta}(r)$ into Eq.~\eqref{eq:Compaction} 
 for $k_*=0.1k_0$, $k_0$, $1.5k_0$ and $2k_0$ are shown in Fig.~\ref{fig:hatzeta}. 
The value of $\bar r_{\rm m}$ is analytically calculated as 
\begin{equation}
\bar r_{\rm m}(k_*)=\sqrt{3} \sqrt{\frac{-3 k_*^2/k_0^2+2+\sqrt{5 k_*^4/k_0^4-8
   k_*^2/k_0^2+4}}{1-k_*^2/k_0^2}}, 
\end{equation}
and is depicted as a function of $k_*$ in Fig.~\ref{fig:rm}. 
%
The functional form of $\mu_{\rm th}^{(k_*)}$ defined in Eq.~\eqref{eq:muth} and $k_*=k_*^{\rm th}(M)$ 
defined as the inverse function of $M=M^{(\mu,k_*)}(\mu_{\rm th}^{(k_*)}(k_*),k_*)$ is 
depicted in Fig.~\ref{fig:muthk}. 
From Fig.~\ref{fig:muthk}, it is clear that $M^{(\mu,k_*)}(\mu_{\rm th}^{(k_*)}(k_*),k_*)$
takes the maximum value $M\simeq 19/k_0$ at $k_*=0$. 
In general, $k_*$ can be regarded as a function of $\mu$ and $M$. 
The behavior of $k_*$ as a function of $M$ for each value of $\mu$ is depicted in Fig.~\ref{fig:kM}.
As is shown in Fig.~\ref{fig:kM}, the maximum value of $M$ is realized at $k_*=0$ for each $\mu$. 
Therefore, for a given value of $M$, the minimum value of $\mu_{\rm min}$ is given by 
\begin{equation}
\mu_{\rm min}(M)=\mu^{(M,k_*)}(M,0).
\end{equation}
Substituting the function $k_*^{\rm th}(M)$ into Eq.~\eqref{eq:muth}, we can draw $\mu_{\rm th}^{(M)}$ as 
a function of $M$ as is shown in Fig.~\ref{fig:muthM}. 
In the small mass region, namely large $k_*$ region, the second term in Eq.~\eqref{eq:zetahat} dominates the 
profile. Then, the magnitude of $k_*$ degenerates with the overall amplitude of the inhomogeneity. 
Consequently, PBH can form even for very small value of $\mu$ if $k_*$ is sufficiently large. 
However, as will be shown below, the probability of such cases are exponentially suppressed. 

\begin{figure}[htbp]
\begin{center}
\includegraphics[scale=0.63]{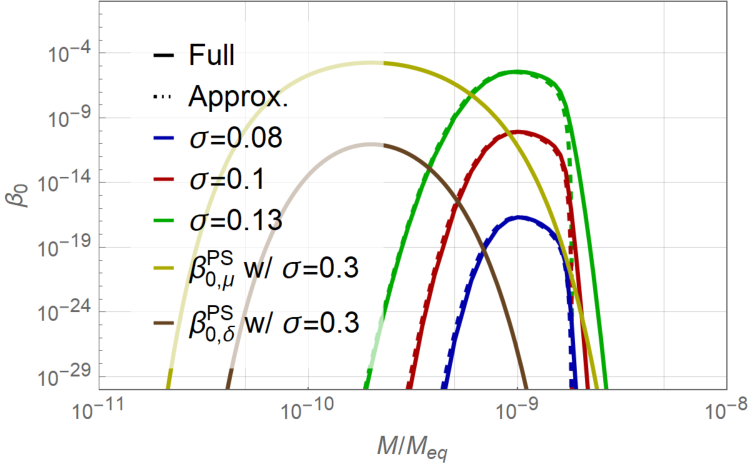}
\includegraphics[scale=0.63]{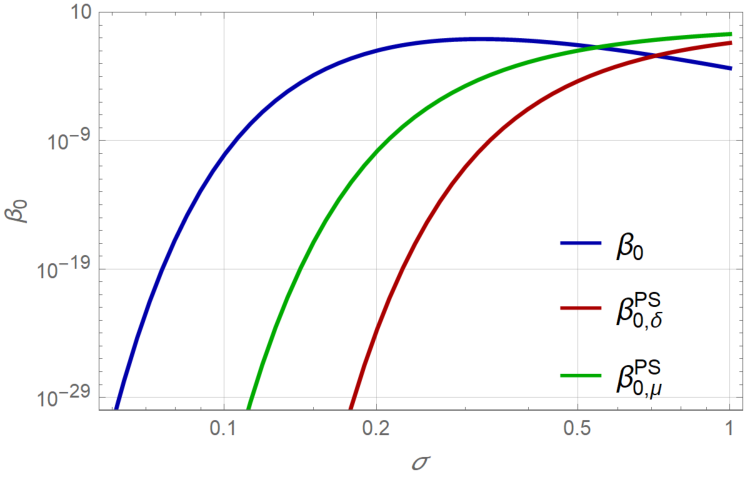}
\caption{
\baselineskip5.5mm
PBH fraction to the total density $\beta_0$ at the formation time 
is depicted as a function of the PBH mass $M$ (left panel) for 
each value of $\sigma_0$, and a function of $\sigma_0$ (right panel) for 
$M=M_{\rm c}$ with $k_0=10^5k_{\rm eq}$ and $\alpha=1$. 
In the right panel, for the PS formalism, the value at the peak of the mass spectrum is taken. 
We also depict the value at the peak of the mass spectrum from the PS formalism  with the Gaussian 
window function in the right panel.  
}
\label{fig:mass_spectrum_ext_1}
\end{center}
\end{figure}

\begin{figure}[htbp]
\begin{center}
\includegraphics[scale=0.63]{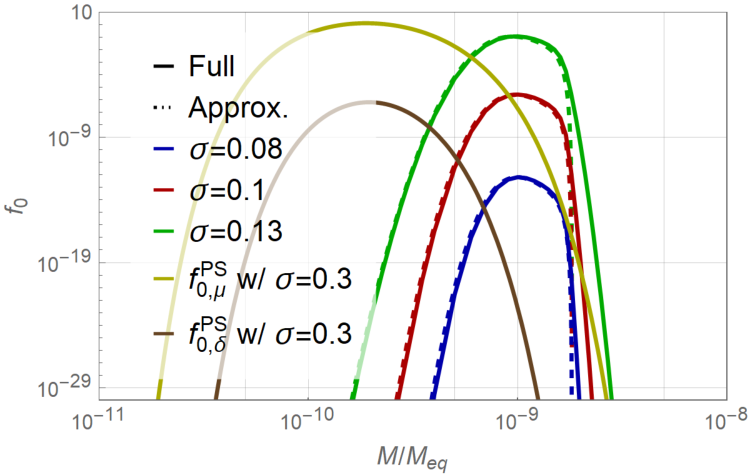}
\includegraphics[scale=0.63]{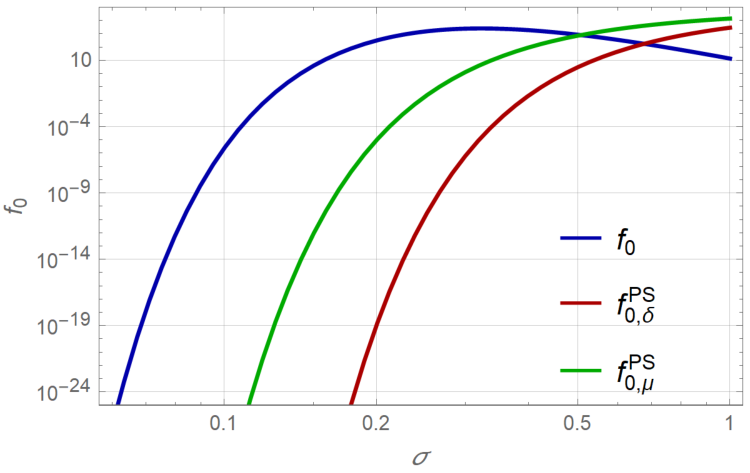}
\caption{
\baselineskip5.5mm
PBH fraction to the total density $f_0$ at the equality time 
is depicted as a function of the PBH mass $M$ (left panel) for 
each value of $\sigma_0$, and a function of $\sigma_0$ (right panel) for 
$M=M_{\rm c}$ with $k_0=10^5k_{\rm eq}$ and $\alpha=1$. 
We also depict the value at the peak of the mass spectrum from the PS formalism  with the Gaussian 
window function in the right panel.  
}
\label{fig:mass_spectrum_ext_2}
\end{center}
\end{figure}

\begin{figure}[htbp]
\begin{center}
\includegraphics[scale=0.7]{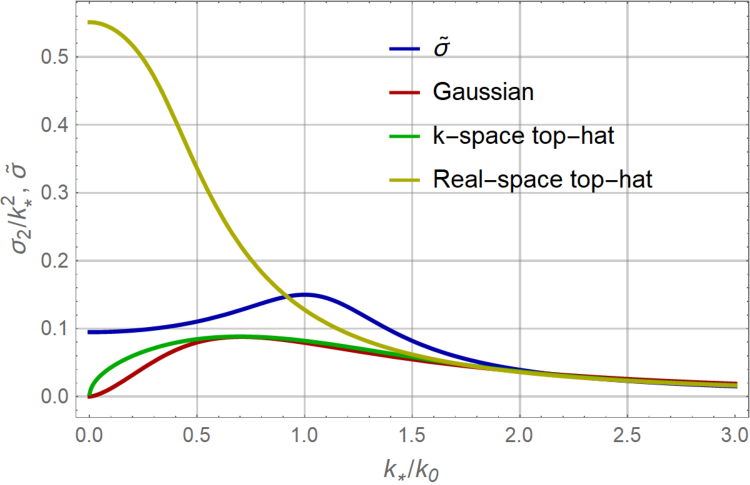}
\caption{$\tilde \sigma$ and $\hat\sigma_2/k_*^2$ as functions of $k_*$. 
}
\label{fig:sigs}
\end{center}
\end{figure}

Using all the functional forms shown above, we can numerically evaluate the integral 
\eqref{eq:beta_general}. 
The results are shown in Figs.~\ref{fig:mass_spectrum_ext_1} and \ref{fig:mass_spectrum_ext_2}. 
%
%
We note that the mass spectrum from the conventional PS formalism is much 
smaller for the same value of $\sigma$. 
There are two reasons for this difference. 
One is the smaller threshold value $0.753<9\delta_{\rm th}/4\approx1.2$ as is also discussed for the monochromatic power spectrum in Sec.~\ref{sec:monochrom}. 
In addition, for the extended spectrum case, we also find that the value of $\beta_{0,\mu}^{\rm PS}(M)\gg\beta_{0,\delta}^{\rm PS}(M)$
is also much smaller than our estimate if we use the following Gaussian window function in the PS formalism: 
\begin{equation}
W_{\rm G}(k/k_*)=\exp\left(-\frac{1}{2}\frac{k^2}{k_*^2}\right).  
\end{equation}
This is because of the difference between the typical dispersions $\hat\sigma_2/k_*^2$ and $\tilde \sigma$. 
In Fig.~\ref{fig:sigs}, we plot $\tilde \sigma$ and $\hat\sigma_2/k_*^2$ for 
the  Gaussian, real-space top-hat, $k$-space top-hat window functions.
The real-space and $k$-space top-hat window functions are defined as follows:
\begin{eqnarray}
W_{\rm RTH}(k/k_*)&=&3\frac{\sin(k/k_*)-k/k_*\cos(k/k_*)}{k^3/k_*^3}, \\
W_{k{\rm TH}}(k/k_*)&=&\Theta(k_*-k). 
\label{eq:windows}
\end{eqnarray}
Fig.~\ref{fig:sigs} shows that the value of $\tilde \sigma$ is larger 
than $\hat \sigma_2/k_*^2$ with the Gaussian window function. 
This difference is reflected to the many orders of magnitude difference in the PBH fraction.

\subsection{Behavior near the spectrum peak}
\label{sec:simple}
In order to obtain a simpler analytic formula, let us consider the typical profile with 
the value of $k_*$ being its mean value $k_{\rm c}$. 
Replacing $k_*^2$ by $k_{\rm c}^2$ in the expression of $g$, we obtain 
\begin{equation}
g(r;k_{\rm c})=-\psi(r). 
\end{equation}
Typical value $g_{\rm c}$ of $g_{\rm m}$ can be given by 
\begin{equation}
g_{\rm c}:=g_{\rm m}(k_{\rm c}). 
\end{equation}
In order to simplify the $k_*$ dependence of $\bar r_{\rm m}$, we assume the following 
simple form 
\begin{equation}
\bar r_{\rm m}(k_*)\approx  \frac{\ell}{k_*}. 
\end{equation} 
Then, from Eq.~\eqref{eq:kM}, the relation between $k_*$ and $M$ can be approximated by 
\begin{equation}
k_*^2\approx k_*^{\rm ap}(M):=k_{\rm eq}^2 \frac{M_{\rm eq}}{M}\ell^2 \ee^{-2\mu_{\rm c}g_{\rm c}}. 
\label{eq:approxkM}
\end{equation}
For Eq.~\eqref{eq:beta_app}, let us perform the following replacement:
\begin{equation}
\bar r_{\rm m}\rightarrow \frac{\ell}{k_*^{\rm ap}},~
\mu_{\rm th}^{(M)}(M)\rightarrow \mu_{\rm c},
~g_{\rm m}(k_*)\rightarrow g_{\rm c}, 
~k_*\rightarrow k_*^{\rm ap}(M). 
\end{equation}
The second replacement implies $\mu_{\rm b}={\rm max}\left\{\mu_{\rm min}(M), \mu_{\rm c}\right\}$. 
In addition, since $\sigma_n\ll1$, we can also approximate the function $f$ as~\cite{1986ApJ...304...15B} 
\begin{equation}
f\left(\frac{\mu_{\rm b} k_*^2}{\sigma_2}\right)\approx \left(\frac{\mu_{\rm b} k_*^2}{\sigma_2}\right)^3. 
\end{equation}
Then, finally, we obtain the following analytic expression:
\begin{equation}
\beta_0^{\rm simp}(M)=\dred{\frac{2 \alpha \ell^8 \mu_{\rm b}^3 \ee^{-8\mu_{\rm b}g_{\rm c}}}{3^{5/2}(2\pi)^{3/2}\sqrt{1-\gamma^2}}}
k_{\rm eq}^5 \left(\frac{M_{\rm eq}}{M}\right)^{5/2}\frac{\tilde \sigma_{\rm ap}(M)^2}{\sigma_0\sigma_1^3\sigma_2}
\exp\left[-\frac{1}{2}\frac{\mu_{\rm b}^2}{\tilde \sigma_{\rm ap}(M)^2}\right], 
\label{eq:simplebeta}
\end{equation}
where 
\begin{equation}
\frac{1}{\tilde\sigma_{\rm ap}(M)^2}=\frac{1}{\sigma_0^2}+\frac{\left({k_*^{\rm ap}(M)}^2-k_{\rm c}^2\right)^2}{(1-\gamma^2)\sigma_2^2}. 
\end{equation}
It should be noted that, while the same peak value as Eq.~\eqref{eq:ext_approx} is given by the above simple formula, 
the curvature of the spectrum peak cannot be well approximated due to the non-trivial scale dependence of 
$\mu_{\rm th}^{(k_*)}$(see Fig.~\ref{fig:muthk}) which is not taken into account in Eq.~\eqref{eq:simplebeta}. 

In order to take $k_*$ dependence of $\mu_{\rm th}^{(k_*)}$ into account, 
let us consider the Taylor expansion of $\mu_{\rm th}^{(k_*)}$ around $k_*=k_{\rm c}$. 
First, we focus on the function $m(r,k_*^2)$ defined by 
\begin{equation}
m(r,k_*^2):=rg'(r;k_*). 
\end{equation}
This function satisfies 
\begin{equation}
\frac{\del }{\del r}m(\bar r_{\rm m}(k_*),k_*^2)=0. 
\end{equation} 
Around $k_*^2=k_{\rm c}^2$, we may consider the following expansion of $m(\bar r_{\rm m}(k_*),k_*^2)$:
\begin{equation}
m(\bar r_{\rm m}(k_*),k_*^2)=m_0+m_1(k_*^2-k_{\rm c}^2)+\frac{1}{2}m_2(k_*^2-k_{\rm c}^2)^2
+\mathcal O\left((k_*^2-k_{\rm c}^2)^3\right), 
\end{equation}
where $m_0$, $m_1$ and $m_2$ are, respectively, given by
\begin{eqnarray}
m_0&=&r_{\rm c}g_{\rm c}'=-r_{\rm c}\psi_{\rm c}',\\
m_1&=&\frac{\del}{\del k_*^2}m(r_{\rm c},k_{\rm c}^2)=r_{\rm c}g_{\rm 1c}', \\
m_2&=&\left(\frac{\dd \bar r_{\rm m}}{\dd k_*^2}\right)_{k_*=k_{\rm c}}^2 \frac{\del^2}{\del r^2}m(r_{\rm c},k_{\rm c}^2)
+2\left(\frac{\dd \bar r_{\rm m}}{\dd k_*^2}\right)_{k_*=k_{\rm c}}\frac{\del^2}{\del r\del k_*^2}m(r_{\rm c},k_{\rm c}^2)
\cr
&=&\left(\frac{\dd \bar r_{\rm m}}{\dd k_*^2}\right)_{k_*=k_{\rm c}}^2(2\psi''_{\rm c}+r_{\rm c}\psi'''_{\rm c})
+2\left(\frac{\dd \bar r_{\rm m}}{\dd k_*^2}\right)_{k_*=k_{\rm c}}(g'_{\rm 1c}+r_{\rm c}g''_{\rm 1c}) 
\end{eqnarray}
with the subscript ``${\rm c}$" denoting the value at $k_*=k_{\rm c}$. 

\begin{figure}[htbp]
\begin{center}
\includegraphics[scale=0.8]{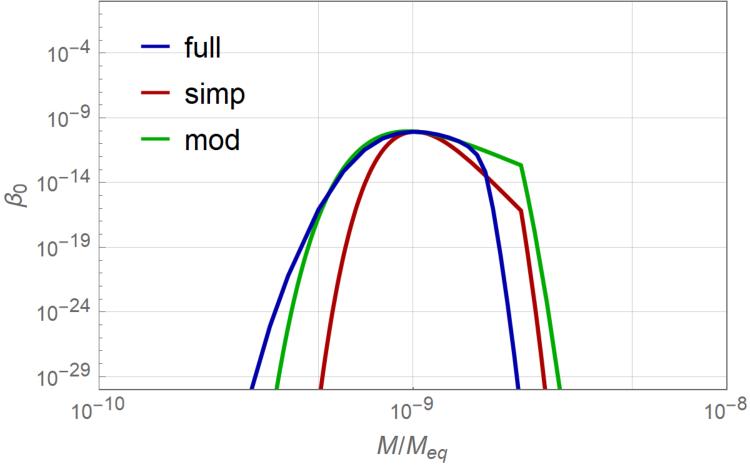}
\caption{$\beta_0$, $\beta_0^{\rm simp}$ and $\beta_0^{\rm mod}$ for $\sigma=0.1$. 
}
\label{fig:simp}
\end{center}
\end{figure}

Then, the threshold value $\mu^{(k_*)}$ is given by 
\begin{equation}
\mu^{(k_*)}(k_*)=\mu_{\rm c}\left(1-\frac{m_1}{m_0}\left(k_*^2-k_{\rm c}^2\right)+\left(\frac{m_1^2}{m_0^2}-\frac{m_2}{2m_0}\right)\left(k_*^2-k_{\rm c}^2\right)^2\right)+\mathcal O((k_*^2-k_{\rm c}^2)^3). 
\end{equation}
Substituting this expression into the combination $\mu^2/\tilde \sigma^2$ in the exponential of the probability \eqref{eq:p1mod}, 
we find
\begin{eqnarray}
\frac{{\mu^{(k_*)}}^2}{\mu_{\rm c}^2 }\frac{\sigma_0^2}{\tilde \sigma^2}&=&1
-2\frac{m_1}{m_0}(k_*^2-k_{\rm c}^2)
\cr
&&+\left(\frac{\sigma_0^2}{(1-\gamma^2)\sigma_2^2}+\frac{2m_1^2}{m_0^2}-\frac{m_2}{m_0}\right)
\left(k_*^2-k_{\rm c}^2\right)^2+\mathcal O((k_*^2-k_{\rm c}^2)^3). 
\end{eqnarray}
For our specific example, 
given by Eq.~\eqref{eq:Gausspower}, we find $m_0=m_2/k_0^4=2/\ee$ and $m_1=0$, and 
the first term in the quadratic coefficient is $3/(2k_0^4)$ 
while the remaining correction terms are $-1/k_0^4$. 
Therefore, the value of the effective variance increases and the spectrum peak is flatten. 
For our specific example, we obtain $m_1=0$. 
However, in general, we may expect non-vanishing value of $m_1$ and it may cause the shift of the spectrum peak. 
Defining the modified dispersion $\tilde \sigma_{\rm mod}$ as 
\begin{equation}
\frac{1}{\tilde \sigma_{\rm mod}(M)^2}:=\frac{1}{\sigma_0^2}\left[1
-2\frac{m_1}{m_0}(k_*^{\rm ap}(M)^2-k_{\rm c}^2)+\left(\frac{\sigma_0^2}{(1-\gamma^2)\sigma_2^2}+\frac{2m_1^2}{m_0^2}-\frac{m_2}{m_0}\right)\right], 
\end{equation}
we obtain the following modified simple expression:
\begin{equation}
\beta_0^{\rm mod}(M)=\dred{\frac{2 \alpha \ell^8 \mu_{\rm b}^3 \ee^{-8\mu_{\rm b}g_{\rm c}}}{3^{5/2} (2\pi)^{3/2}\sqrt{1-\gamma^2}}}
k_{\rm eq}^5 \left(\frac{M_{\rm eq}}{M}\right)^{5/2}\frac{\tilde \sigma_{\rm mod}(M)^2}{\sigma_0\sigma_1^3\sigma_2}
\exp\left[-\frac{1}{2}\frac{\mu_{\rm b}^2}{\tilde \sigma_{\rm mod}(M)^2}\right]. 
\label{eq:simple_mod_beta}
\end{equation}
This modified version of the simple expression gives better approximation for the shape of the spectrum around the peak as is shown in Fig.~\ref{fig:simp}. 
%

\section{Summary and Discussion}

Primordial black holes (PBHs) have attracted much attention not
only from a theoretical point of view but also observationally.  In order to
make an observationally relevant prediction from a theoretical result,
the estimation of the abundance of PBHs is essential.  

The conventional Press-Schechter(PS) formalism assumes a Gaussian distribution of the primordial density or curvature perturbation.
However, the threshold value of the curvature perturbation has an
ambiguity from environmental effects~\cite{Young:2014ana,Harada:2015yda}. On the other hand, the existence of an
upper limit for the density perturbation
at the time of horizon entry~\cite{Kopp:2010sh}
 is in conflict with the naive assumption of a Gaussian probability distribution for such variable.

In order to overcome the above issues, we have developed a formalism
to estimate the abundance of PBHs by combining the Gaussian probability
distribution of the curvature perturbation with a threshold value for
the locally averaged density perturbation. More precisely,
we consider an optimized criterion for PBH formation which
is given by a threshold for the compaction function proposed in Ref.~\cite{Shibata:1999zs}. 
The compaction function is given by the value of the mass excess above the background 
divided by the area radius.  
Our approach is based on the peak theory of Gaussian random fields~\cite{1986ApJ...304...15B}, and takes into consideration the non-linear
relation between the curvature perturbation and the density perturbation.  
A general expression
for the fraction of PBHs to the total density of the Universe is presented. 
Although our current procedure is only directly applicable to 
a spectrum which has a localized peak around a specific scale, 
the expression for the PBH fraction, in principle, can be evaluated once the curvature power spectrum is given. 
In our program to calculate the PBH mass spectrum, there is no need to introduce a window function
as far as we consider a power spectrum localized around a single scale in the Fourier space. 
The scale dependence of the PBH fraction, which is conventionally induced by a window function, 
is induced by considering a typical peak profile of the curvature perturbation 
characterized by the amplitude($\mu$) and the spatial peak curvature($k_*^2$) 
for a given curvature power spectrum. 
We note that, for each profile characterized by $\mu$ and $k_*$, at the horizon crossing time, 
the PBH mass is evaluated 
from the mass inside the radius at which the compaction function 
takes the maximum value. 

The case of the monochromatic power spectrum is particularly simple, and illustrative of the
general case. 
First of all, compared to the conventional PS approach, the PBH spectrum is shifted to larger masses, 
by an order of magnitude or so. 
This is due to the 
%
mass estimation from the mass inside the maximum compaction radius,  
which is substantially affected by the metric perturbation. 
A related effect is that there is a slight spread in the mass spectrum, even when the
underlying primordial spectrum is monochromatic. Such spread, however, is hardly significant.
Finally, the estimated abundance of PBH is much larger than in the conventional
PS formalism, by many orders of magnitude. This effect comes mainly from the optimized PBH 
formation criterion. Roughly speaking, in the limit when the probability of PBH formation is exponentially low, our 
estimate for such probability is larger than the fourth root of the conventional result. Therefore, the effect is very significant.
We have also considered the case of an extended power spectrum. 
For the extended power spectrum, we also have qualitatively similar effects. 
In addition to those effects, we found that the PBH fraction is yet larger than the value given by the PS formalism with the Gaussian window function. 
This is simply because there is no unphysical 
suppression due to the window function in our formalism.

Throughout this paper we have assumed that the criterion for the formation of PBHs
can be given in terms of a threshold for the amplitude
of some (averaged) density perturbation.
However, it is known that such threshold depends on the profile of the overdensity.  
In order to clarify
such dependence, we need an analysis combined with numerical simulations of the gravitational collapse. 
Another interesting aspect to consider is the effect of a non-Gaussianity in the primordial curvature
perturbation. 
In this paper, we have assumed for simplicity that the primordial 
curvature perturbation is given as a Gaussian random field. 
Note that, even in this case, non-Gaussian features are generally induced by 
non-linear relations between perturbation variables. 
Primordial non-Gaussianity in single field models of inflation is usually suppressed 
by the slow roll parameters, but even so its effect on PBH formation 
could be significant. 
Furthermore, the generation of a bump in the power 
spectrum typically involves a short period of ``ultra-slow roll", which 
may entail additional non-Gaussianity near the scale of the bump 
responsible for black hole formation (note that such non-Gaussianity is 
on small scales, and therefore it is currently very poorly constrained by 
CMB data). 
The degree of such non-Gaussianity is model dependent, and 
the corresponding enhancement/suppression of the abundance of PBHs and 
clustering of PBHs due to the non-Gaussianity are currently under active 
investigation~\cite{Franciolini:2018vbk}. 
The incorporation of primordial non-Gaussianity would 
require an extension of the procedure presented here. 
Such issues are left 
for further research. 
%

\section*{Acknowledgements}
We thank T. Hiramatsu, I. Musco, M. Sasaki, T. Suyama, T. Takeuchi and T. Tanaka
for helpful comments.  This work was supported by JSPS KAKENHI Grant
Numbers JP16K17688, JP16H01097 (C.Y.), JP26400282 (T.H.), JP17H01131
(K.K.), 26247042 (K.K.), and MEXT KAKENHI Grant Numbers
JP15H05889(K.K.), and JP16H0877 (K.K.) FPA2016-76005-C2-2-P, MDM-2014-0369 of ICCUB 
(Unidad de Excelencia Maria de Maeztu), AGAUR 2014-SGR-1474 (J.G.).The authors thank the Yukawa
Institute for Theoretical Physics at Kyoto University.  Discussions
during YITP-T-17-02 ``Gravity and Cosmology 2018" and the YITP
symposium YKIS2018a ``General Relativity -- The Next Generation --"
were useful to complete this work.

\appendix

\section{Deviation between peaks of $\zeta$ and $\delta$}
\label{devpeaks}
Our analysis is based on the Taylor expansion around a peak of
$\zeta$.  However, since the criterion for a formation of PBHs is
given in terms of $\delta$, one may concern about the deviation
between each peak of $\zeta$ and $\delta$. 
First, let us define the renormalized local scale factor $\bar a=a\ee^{-\zeta_0}$.
Assuming the all
moments $\sigma_n$ are far smaller than $k_*^n\sim(\bar a H)^{n}$ at
the horizon entry, we show that, if the value of $\delta$ is
comparable to the threshold value $\delta_{\rm th}$ at a peak, we can
almost always find the associated peak of $\zeta$ well inside the
horizon patch centered at the peak of $\delta$.  This fact means that
the region which allows the formation of PBHs typically involves peaks
of $\delta$ and $\zeta$ near the center, and validates our procedure.

Let us consider a peak of $\delta$ (not peak of $\zeta$) satisfying
$\delta>\delta_{\rm th}$.  The value of $\delta$ can be expressed in
terms of the Taylor expansion around the peak as follows:
\begin{equation}
\delta=\frac{2(1+w)}{3w+5}\frac{1}{a^2H^2}\left[\ee^{2\zeta_0}\left(\zeta_2-\frac{1}{2}\sum_i
\left(\zeta_1^i\right)^2\right)\right]+\mathcal O(y^2), 
\end{equation}
where we introduced a new spatial coordinate $\bm y$ to emphasize
the difference from the expansion around the peak of $\zeta$ adopted
in the text.  
From the inequality $\delta>\delta_{\rm th}$, we obtain
the following inequality:
\begin{equation}
\zeta_2>\frac{3w+5}{2(1+w)}\bar a^2 H^2\delta_{\rm th}\gg \sigma_2,  
\end{equation}
where
and hereafter we evaluate every variable at the horizon entry,
that is, $k_*\sim\bar a H$.  The deviation $\Delta y_j$ of the peak of
$\zeta$ from the peak of $\delta$ can be estimated by the following
equation:
\begin{equation}
\sum_j \Delta y_j \zeta_2^{ji}=-\zeta_1^i. 
\end{equation}
Taking the principal direction of $\zeta_2^{ij}$, we obtain 
\begin{equation}
\zeta_2^{ij}={\rm diag}\left\{\lambda_1,\lambda_2,\lambda_3\right\}. 
\end{equation}
Since $\zeta_2\gg \sigma_2$, from the probability distribution of
$P_2(\xi_2,\xi_3,\bm \eta)$, we can find that the probability
to have a negative eigen value is typically very small.  Then, we
obtain the following equation:
\begin{equation}
|\Delta y_i|=|\zeta_1^i/\lambda_i|\lesssim \sigma_1 /k_*^2\ll 1/(\bar a H),  
\end{equation}
where we have used $\lambda_i \sim \zeta_2\gtrsim k_*^2$ and $\zeta_1^i\sim\sqrt{<\zeta^i_1\zeta^i_1>}\sim\sigma_1$.

\section{Correspondence with 3-zone model}
Let us consider the corresponding parameters to $\zeta_0$, $\mu$ and $k_*$ 
in 3-zone model\cite{Harada:2013epa}  according to Refs.~\cite{Harada:2013epa,Harada:2015yda}. 
In the 3-zone model, we consider the closed FLRW model as the collapsing region. 
The line element can be written as the following forms:
\begin{eqnarray}
\dd s_3^2&=&a^2\left(\frac{\dd \bar r^2}{1-\bar r^2}+\bar r^2 \dd\Omega^2\right)\\
&=&a^2\ee^{-2\zeta}\left(\dd r^2+r^2 \dd \Omega^2\right). 
\end{eqnarray}
From the metric forms, we obtain the following relations:
\begin{eqnarray}
\frac{\dd \bar r}{\sqrt{1-\bar r^2}}&=&\pm \ee^{-\zeta} \dd r, \\
\frac{\bar r}{r}&=&\ee^{-\zeta}. 
\end{eqnarray}
Then, we obtain 
\begin{equation}
\frac{r}{C}=\left\{
\begin{array}{lll}
\frac{2\bar r}{1+\sqrt{1-\bar r^2}}&{\rm for}&r\leq 2C\\
\frac{2\left(1+\sqrt{1-\bar r^2}\right)}{\bar r}&{\rm for}&2C<r
\end{array}
\right., 
\end{equation}
\begin{equation}
\bar r=\frac{4r/C}{4+\left(r/C\right)^2} 
\end{equation}
and
\begin{equation}
\ee^{-\zeta}=\frac{\bar r}{r}=\frac{4/C}{4+\left(r/C\right)^2},  
\end{equation}
where $C$ is the integration constant. 
This constant is related to $\zeta_0$ as $\ee^{-2\zeta_0}=1/C^2$ and can be absorbed into 
the renormalized scale factor $\bar a$ as in the text or equivalently into 
the rescaling of the radial coordinate. 
Therefore, hereafter, we set $C=1$ for simplicity.

The value of $\triangle \zeta$ at the origin can be evaluated as 
\begin{equation}
\left.\triangle\zeta\right|_{\bar r=0}=\frac{3}{2}=\mu k_*^2, 
\end{equation}
where we have identified it as $\mu k_*^2$ as in the text. 
From the Hubble equation, the density perturbation $\delta^{\rm CMC}$ in the uniform Hubble slice is given by 
\begin{equation}
\delta^{\rm CMC}=\frac{8\pi \rho}{3H^2}-1=\frac{1}{a^2H^2}. 
\end{equation}

Let us define the radius $r_*$ corresponding to 
the scale $1/k_*$ as 
\begin{equation}
r_*^2=q/k_*^2=\frac{2}{3}q\mu,  
\end{equation}
with $q$ being a numerical factor. 
Then, we identify the horizon entry by 
\begin{equation}
\frac{1}{aH}=\bar r_*, 
\label{hentcondApp}
\end{equation}
where 
\begin{equation}
\bar r_*:=\frac{4r_*}{4+r_*^2}. 
\end{equation}
Then, $\delta^{\rm CMC}$ at the horizon entry time $\delta^{\rm CMC}_{\rm H}$ is given by 
\begin{equation}
\delta^{\rm CMC}_{\rm H}=\frac{24q\mu}{(6+q\mu)^2}. 
\end{equation}
This quantity takes the maximum value 1 at $\mu=6/q$. 
For a small radius, we may approximate $r_*\simeq \bar r_*$, and 
comparing the horizon entry conditions \eqref{hentcondApp} and \eqref{hentry}, 
we can estimate the value of $q$ as $\sim 10$(see Table\ref{tab:examples} in Appendix\ref{sec:exs}). 
%
This relation between $\delta^{\rm CMC}_{\rm H}$ and $\mu$ explicitly shows 
that the value of the density perturbation at the horizon entry time has a maximum value 
while the value of $\mu$ is not bounded. 
The case $\mu=6/q$ corresponds to the 3-hemisphere which divides the Type I and Type II PBH formation. 

\section{examples of the curvature profile} \label{sec:exs}
Let us focus on the case in which the curvature profile $g(r;k_*)$ is given by 
a function of $k_*r$ 
with the characteristic scale $1/k_*$. 
Then, for notational simplicity, we denote $g(r;k_*)$ as $g(k_* r)$ in this section.   
Here we list some examples of the function $g(x)$ with $g(0)=-1$ and $g''(0)=1/3$. 
\begin{itemize}
\item{Sinc function}
\begin{equation}
g(x)=-{\rm sinc}(x)=\sum_{n=0}^\infty\left[\frac{(-1)^{n+1}}{(2n+1)!}x^{2n}\right]. 
\label{eq:sinc}
\end{equation}
\item{up to 4th order}
\begin{equation}
g(x)=\sum_{n=0}^2\left[\frac{(-1)^{n+1}}{(2n+1)!}x^{2n}\right]=-1+\frac{1}{6}x^2-\frac{1}{120}x^4. 
\end{equation}
\item{up to 8th order}
\begin{equation}
g(x)=\sum_{n=0}^4\left[\frac{(-1)^{n+1}}{(2n+1)!}x^{2n}\right]. 
\end{equation}
\item{Gaussian}
\begin{equation}
g(x)=-\exp\left(-\frac{x^2}{6}\right).  
\label{eq:gauss}
\end{equation}
\item{Dawson function}
\begin{equation}
g(x)=-\frac{2D_+(k_*r/2)}{k_* r}. .  
\end{equation}
\end{itemize}
The values of $\ell:=r_{\rm m}(k_{\rm c})k_{\rm c}$, $-g(\ell)$, $\mu_{\rm th}:=(2-\sqrt{4-6\delta_{\rm th}})/(2\ell g'(\ell))$ and $\ell^2 \ee^{-2\mu_{\rm th}g(\ell)}$ are listed in Table \ref{tab:examples}. According to 
Eq. (\ref{eq:kM}), the latter column represents the enhancement in the black hole mass relative to the naive linear estimate. Note that this effect is of a similar magnitude in all five examples.
\begin{table}[htbp]
\caption{The values of $\ell$, $g(\ell)$, $\mu_{\rm th}$ and $\ell^2 \ee^{-2\mu_{\rm th}g(\ell)}$
for the listed examples with $\delta_{\rm th}\approx 0.533$. 
}
\label{tab:examples}
\begin{tabularx}{160mm}{cCCCC}
&$\ell$&$-g(\ell)$&$\mu_{\rm th}$&$\ell^2 \ee^{-2\mu_{\rm th}g(\ell)}$\\
\hline
\hline
sinc&$2.74$&$0.141$&$0.520$&$8.72$\\
4th order&$\sqrt{5}\simeq2.24$&$3/8=0.375$&$0.663$&$8.22$\\
8th order&$2.72$&$0.151$&$0.522$&$8.66$\\
Gaussian&$\sqrt{6}\simeq2.45$&$1/\ee\simeq0.368$&$0.753$&$10.4$\\
Dawson&$2.35$&$0.436$&$0.865$&$11.7$\\
\hline
\end{tabularx}
\end{table}

\section{Relation between $k_*$ and $k_\delta$}
\label{sec:k-relation}
The linear relation $\delta\sim\triangle \zeta$ suggests that $k_*$ is similar to the inverse length-scale of the density inhomogeneity $k_\delta\equiv (-\triangle \delta/\delta|_{{\bf x}=0})^{1/2}$.
Here, we consider the non-linear relation between $k_\delta$ and $k_*$. 
First, we define $\mu_4$ as follows:
\begin{equation}
\mu_4=-\left.\frac{\triangle \triangle \zeta}{k_*^4}\right|_{\bm x=0}. 
\label{eq:defmu4}
\end{equation}
Using the non-linear relation between $\zeta$ and $\delta$, 
and assuming $\left.\del_i \zeta\right|_{\bm x=0}=0$, 
we have 
$$
k_{\delta}^2 = \left(\frac{\mu_4}{\mu}-\mu\left[2-\frac{\zeta_2^{ij} \zeta_2^{ij}}{\zeta_2^2}\right]\right) k_*^2,
$$
where summation of repeated indices is understood. 
Noting that $\zeta_2^2\geq\zeta_2^{ij} \zeta_2^{ij}$, 
it is clear that $k_*$ and $k_\delta$ will be similar for $\mu_4 \sim \mu \ll 1$. 
However, the case of interest to us is 
$\mu > \mu_{\rm th} \sim 0.5 - 0.8$. 
These are high peaks of the random distribution, and consequently they are nearly spherical. 
In this case, we 
can estimate $k_{\delta}^2 \approx (\mu_4/\mu-5\mu /3) k_*^2$, which becomes negative for $\mu^2>3\mu_4/5$,  
where the density profile has a local minimum at the center of the overdensity.
The typical value of $\mu_4$ is given by substituting $\bar \zeta$ into Eq.~\eqref{eq:defmu4}. 
For instance, we obtain $\mu_4\sim\mu$ for the sinc function \eqref{eq:sinc} and 
$5\mu/3$ for the Gaussian function \eqref{eq:gauss}. 
Then the inequalities for the realization of a local minimum of the density profile 
at the center are given by 
$\mu>3/5$ for the sinc function \eqref{eq:sinc} 
and $\mu>1$ for the Gaussian function \eqref{eq:gauss}.
Since $\mu_{\rm th}$ is smaller than these values, we do not have to worry about this peculiar 
regime, 
where the density profile has a local minimum at the center of the overdensity.\footnote{
  \baselineskip5mm
  After the publication of this paper, it has been shown that, through numerical analyses of the case for sinc function,
the threshold value of $\mu$ is given by 0.61 slightly larger than 3/5~\cite{Atal:2019erb}.
}


\end{document}